\documentclass[a4paper,11pt]{article}
\pdfoutput=1 

\usepackage{jheppub} 
\usepackage{bm}
\usepackage{tikz}
\usepackage{multirow}
\usepackage{amsmath}
\usepackage{slashed}
\usepackage{braket}
\usepackage{amstext,amssymb}
\usepackage{graphicx}

\usepackage{xspace}
\usepackage{color}
\usepackage{units}

\newcommand{\mathsym}[1]{{}}

\newcommand{\hs}{\hspace*{0.02 true cm}}

\newcommand{\bav}{\begin{array}{cccc}}

\usepackage{wrapfig}
\definecolor{light-gray}{gray}{0.95}
\usepackage{tcolorbox}

\begin{document} 

\vspace{0.2cm}

\begin{center}
{\Large\bf Analysis of neutrino oscillation parameters in the light on quantum entanglement}
\end{center}

\vspace{0.1cm}

\begin{center}

 {Rajrupa Banerjee$^\dagger$}~\footnote{rajrupab@iitbhilai.ac.in}, {Papia Panda$^\ddagger$}~\footnote{ppapia93@gmail.com}, {Rukmani Mohanta$^\ddagger$}~\footnote{rmsp@uohyd.ac.in}, {Sudhanwa Patra$^{\dagger,*}$}~\footnote{sudhanwa@iitbhilai.ac.in}
\\
\vspace{0.2cm}
{\it $^\dagger$ Department of Physics, Indian Institute of Technology Bhilai, Durg-491002, India\\
\it $^\ddagger$School of Physics, 
University of Hyderabad, Hyderabad, India-500046\\
\it $^*$Institute of Physics, Bhubaneswar, Sachivalaya Marg, Bhubaneswar 751005, India}
\end{center}
\begin{abstract}
 NNumerous neutrino experiments have confirmed the phenomenon of neutrino oscillation, providing direct evidence of the quantum mechanical nature of neutrinos. In this work, we investigate the entanglement properties of neutrino flavor states within the framework of three-flavor neutrino oscillation using two major entanglement measures: entanglement of formation (EOF) and concurrence, utilizing the DUNE experimental setup. 
Our findings indicate that the maximally entangled state appears between $\nu_{\mu}$ and $\nu_{\tau}$ whereas, $\nu_{e}$ behaves as a nearly separable state.
To further explore the nature of bipartite entanglement, we introduce the concept of the monogamy of entanglement, which allows us to investigate the distinction between genuine three-flavor entanglement and bipartite entanglement. Our analysis confirms that the three-flavor neutrino system forms a bipartite entanglement structure, adhering to the Coffman-Kundu-Wootters (CKW) inequality. Additionally, we implement a minimization procedure to find the best-fit values of the oscillation parameters that correspond to the concurrence minima at the two specific energy points where the concurrence reaches its lowest values. Using these best-fit values, we probe three fundamental unknowns in neutrino oscillation: CP violation sensitivity, neutrino mass hierarchy, and the octant issue of $\theta_{23}$, across two distinct energy points. Our results manifest that while the best-fit values obtained through concurrence minimization show slightly reduced
sensitivity to CP violation compared to current best-fit values, they exhibit greater sensitivity to the mass hierarchy. Furthermore, the study reveals a maximal mixing angle for the atmospheric sector.
\end{abstract}

\def\thefootnote{\arabic{footnote}}
\setcounter{footnote}{0}

\newpage

\section{Introduction}
Neutrino oscillation is an intrinsic quantum mechanical phenomenon that arises directly from the superposition principle applied to two distinct bases of neutrinos: the mass basis and the flavor basis. This phenomenon results in the mixing of the three neutrino flavor states during propagation, as these flavor states are coherent superpositions of mass eigenstates \cite{Bilenky:2004xm,Kayser:2012ce, SNO:2002tuh, T2K:2011ypd, MINOS:2011amj,K2K:2006yov,DayaBay:2012fng, Super-Kamiokande:2002ujc, NOvA:2016vij}. The occurrence of neutrino oscillation is a direct consequence of neutrinos having non-zero mass. The phenomenon of neutrino oscillation is governed by several key parameters: two mass square differences $\Delta m_{21}^{2}$ and $\Delta m^{2}_{31}$, three mixing angles, $\theta_{12},\theta_{23}$ and $\theta_{13}$, charge-parity (CP) violating phase $\delta_{CP}$, baseline length or the distance between the source and detector ($L$), energy of the neutrino ($E$) and the matter term $A=2\sqrt{2}G_{F} N_e E$, where $N_{e}$ is the electron number density and $G_{F}$ is the Fermi constant. Recent experimental data have provided precise values for the mixing angles $\theta_{12}$ and $\theta_{13}$, solar mass difference $\Delta m_{21}^2$,  however, the exact value of the third mixing angle, $\theta_{23}$, remains unknown, and determining the octant of $\theta_{23}$ (whether it is greater or less than $45^\circ$) is a critical challenge in understanding neutrino oscillation. In addition, two major unresolved issues in neutrino physics are the mass hierarchy problem, i.e., whether the $\nu_{3}$ mass eigenstate is heavier (normal hierarchy) or lighter (inverted hierarchy) than the other two mass eigenstates $\nu_{1}$ and $\nu_{2}$, and the determination of the CP violating phase $\delta_{CP}$. The uncertainties in these parameters present significant obstacles to a comprehensive understanding of neutrino interactions and the complete picture of neutrino oscillation.

Quantum entanglement, a fundamental consequence of the superposition principle, is a defining characteristic of quantum systems and serves as the cornerstone of quantum information science \cite{Nielsen_Chuang_2010, PhysRevA.98.052351}. This field underlies practical applications such as quantum teleportation \cite{PhysRevLett.70.1895} and quantum cryptography \cite{PhysRevLett.67.661}, rooted in the concept of non-local correlations, where the sharing of entangled states among the components of a composite system enables these technologies. Several methods exist for quantifying quantum entanglement, with Entanglement of Formation (EOF) \cite{PhysRevLett.70.1895, PhysRevA.54.3824} and concurrence \cite{PhysRevLett.80.2245,PhysRevA.101.032301} being among the most prominent. These measures are crucial for determining the separability condition of an entangled state. EOF quantifies the entanglement shared among the components of a composite system and is defined through the von Neumann entropy of the reduced density matrix of the subsystems, with a zero value indicating a separable state \cite{Mendes-Santos_2020}. However, the logarithmic nature of EOF presents challenges in precisely quantifying shared entanglement. Concurrence, another widely utilized measure, offers a more tractable approach compared to EOF. Additionally, Monogamy of Entanglement (MOE) serves as a fundamental framework \cite{Zong:2022xza} for understanding the distribution of quantum correlations in multipartite systems, particularly in justifying the non-local nature of these correlations \cite{Henderson:2001wrr} and the constraints on the shareability of entanglement among different subsystems \cite{Coffman_2000, D_r_2000}. It provides a criterion for bipartite separability within a larger multipartite context \cite{Sk:2021dbf, Swain:2021btc}. Coffman, Kundu, and Wootters introduced a simplified formulation of this concept through the CKW inequality, offering a quantitative approach to capture the constraints imposed by monogamy of entanglement in multipartite entangled systems \cite{Coffman:1999jd}.

Given that both neutrino oscillation and quantum entanglement originate from the quantum superposition principle,  investigation of potential connections between these phenomena presents a compelling research avenue \cite{Blasone:2007vw, Blasone:2010ta, Blasone:2013zaa, Blasone:2014jea, Blasone:2015lya, Banerjee:2015mha}. Extensive theoretical work has laid a foundational understanding of entangled states in neutrinos, particularly within the flavor basis \cite{Ettefaghi:2023zsh, Blasone:2022joq, Li:2022mus, Ettefaghi:2020otb}. 
In the neutrino sector, it is a well-established concept that flavor eigenstates, corresponding to the interaction eigenstates, are quantum superpositions of mass eigenstates, also known as propagation eigenstates. This relationship is particularly evident at the moment of neutrino production, where the neutrino exists in a pure quantum state.
During propagation, however, the interaction between the flavor and mass bases results in a dynamic interchange, wherein the mass eigenstates acquire flavor compositions and vice versa. Consequently, the flavor eigenstates evolve into superpositions of different flavor components \cite{Smirnov:2003da}. Notably, this superposition is not a simple quantum superposition but may instead give rise to an entangled superposition state \cite{Blasone:2007vw, Blasone:2007wp}. The fundamental distinction between a simple quantum superposition and an entangled superposition lies in the fact that the latter cannot be factorized into a direct product of two independent component states.

Studies have demonstrated the existence of entangled states in both bipartite and tripartite configurations of neutrino flavor states from a theoretical perspective \cite{Bittencourt:2023asd, Mohammed:2019svq, Floerchinger:2018mla, Akhmedov:2010ua, Kayser:2010bj}. This work aims to investigate the possibility of the existence of such entangled superposition states in the context of neutrino oscillations, providing new insights into their quantum behavior and underlying structure in the experimental aspect. Since the entanglement arises through the dynamical changes of the flavor states, it is reasonable to verify the amount of entanglement in terms of two entanglement-measures such as EOF and concurrence. This is because EOF represents a logarithmic function of these probabilities \cite{Konwar:2024nrd} whereas, concurrence is directly tied to the probabilities of the appearance and disappearance of the three neutrino flavor states \cite{Jha:2020qea, KumarJha:2020pke} during oscillation. 
 
Despite these theoretical advancements, experimentally probing the existence of entangled neutrino states and applying these concepts to address unresolved questions in neutrino physics remain formidable challenges. 
Several theoretical proposals have been advanced to address the unresolved questions in neutrino physics \cite{Blasone:2023qqf, Blasone:2007vw, Quinta:2022sgq}. These approaches aim to explain key issues such as the nature of neutrino mass \cite{, Dixit:2018kev,Naikoo:2017fos}, the mechanism behind neutrino oscillations \cite{Blasone:2014cub, Blasone:2007wp}, and the potential for CP violation in the lepton sector that also include the extensions to the Standard Model in presence of non-standard neutrino interactions \cite{Konwar:2024nrd, Dixit:2020ize}.
While significant progress has been made \cite{Blasone:2023qqf, Boyanovsky:2011xq,Ahluwalia:2011ea, Kayser:2010bj, Li:2021epj}, a comprehensive phenomenological analysis that incorporates these aspects continues to be an open area for further exploration.

In this work, we demonstrate the formation of maximally entangled states within the experimental setup of Deep Underground Neutrino Experiment (DUNE) and extend this to test the entanglement monogamy relation, establishing quantum correlations between flavor states and entanglement shareability in DUNE. Utilizing the entanglement measures EOF and concurrence, we reveal that in the DUNE setup, $\nu_{\mu}$ and $\nu_{\tau}$ form a maximally mixed state that satisfies the monogamy inequality for the three-flavor neutrino oscillation. We employ the Coffman-Kundu-Wootters (CKW) inequality to verify the monogamy relation in the entangled state, which indicates the non-shareability of entanglement between the $\nu_{\mu}$ and $\nu_{\tau}$ flavor states. With this foundational understanding of entangled flavor states in DUNE, we proceed to apply the concept of concurrence minimization to determine a set of best-fit values for the neutrino oscillation parameters in the context of quantum information. These results prompt further investigation into the experiment's sensitivity for addressing three major questions in neutrino physics: the CP violation, mass hierarchy, and octant degeneracy in atmospheric mixing angle.

The structure of this paper is as follows. We begin with the theoretical framework, laying the mathematical foundation of our study. In this section, we introduce the basic concepts of entanglement measures and their expression in the probabilistic context of neutrino oscillation. We also explore the fundamental aspects of EOF, concurrence, and entanglement monogamy. Following this, in Section \ref{exp}, we detail the experimental setup of DUNE and simulation procedures. Section \ref{analysis} is dedicated to the analysis of EOF, concurrence, and the monogamy inequality in DUNE experimental configuration. In the final subsection of this part, we introduce the extremization process for the entanglement measures and determine the best-fit values of the oscillation parameters. Finally, in Section \ref{resultss}, we use these best-fit values from our defined framework to analyze the results concerning CP violation, mass hierarchy, and octant sensitivity. The paper concludes with a summary that encapsulates the main findings and significance of the work.

\section{Theoretical Framework}
\label{theory}
Consider an ultra-relativistic neutrino (left-handed) with flavor $\alpha$ with $\alpha=e,\mu,\tau$ and momentum $\Vec{p}$. The current theoretical framework accounts for two distinct eigenstates of neutrinos: the mass eigenstate, which is the propagating state, and the flavor eigenstate, which is the interacting state. In neutrino oscillations, the flavor state is expressed as a linear superposition of three mass eigenstates, reflecting the dual bases of neutrinos: the mass basis: $\ket{\nu_{k}},(k=1,2,3)$ and flavor basis $\ket{\nu_{\alpha}},(\alpha=e,\mu.\tau)$. These two bases are connected through a unitary transformation $\ket{\nu_{\alpha}} = U\ket{\nu_{k}}$, where $U$ is the neutrino mixing matrix. We work in a basis in which charge leptons are already diagonal, the light neutrino mass matrix is diagonalised by a unitary mixing matrix known as  Pontecorvo, Maki, Nakagawa, Sakata (PMNS) mixing matrix, $U_{\rm PMNS}\equiv U$ \cite{Pontecorvo:1957qd, 10.1143/PTP.28.870, Pontecorvo:1967fh, Patra:2023ltl}
\begin{equation*} 
U =\begin{pmatrix} 
 c_{12} c_{13} & s_{12} c_{13} & s_{13}e^{-i\delta_{CP}}\\
 -s_{12} c_{23} -c_{12} s_{13} s_{23}e^{i\delta_{CP}} & c_{12} c_{23} -s_{12} s_{13}s_{23}e^{i\delta_{CP}} & c_{13} s_{23}\\
 s_{13} s_{23}-c_{12} s_{13} c_{23}e^{i\delta_{CP}}   & -c_{12} s_{23} -s_{12} s_{13}c_{23}e^{i\delta_{CP}}  & c_{23} c_{13}  
 \end{pmatrix},
 \end{equation*}
 where, $c_{ij}= \cos{\theta_{ij}}$ and $s_{ij}=\sin{\theta_{ij}}$ ($i,j=1,2,3$). Neutrino oscillations can be significantly altered in the presence of Earth's matter. This modification arises due to the coherent forward scattering of neutrinos with particles in the medium, which introduces a potential term, $V_{CC}$, given by $V_{CC}=\sqrt{2}G_{F}N_{e}$ \cite{PhysRevD.17.2369,Mikheyev:1985zog, Bethe:1986ej, Smirnov:2004zv}. To analyze three-flavor neutrino oscillations in matter, we employ the density matrix formalism and use the $\alpha-s_{13}$ approximation \cite{Akhmedov_2004, Asano:2011nj} to construct the qubit state of the neutrino. The density matrix formalism provides a robust framework for analyzing quantum systems, capturing both the appearance and disappearance probabilities of neutrinos. The effective Hamiltonian in the flavor basis is then expressed as,
\begin{equation}
   \mathcal{H}=\frac{\Delta m_{31}^{2}}{2E}\left[U \hs\mbox{diag}(0,\alpha,1)\hs U^{\dagger}+\mbox{diag}(\Hat{A},0,0)\right],
\label{eq:2.1}
\end{equation}
where $\alpha=\Delta m_{21}^{2}/\Delta m_{31}^{2}$, $\Hat{A}=A/\Delta m^{2}_{31}$ 
with $A=2 V_{CC}$. 
 It is to be mentioned here that in the three-flavor neutrino oscillation framework, only Charged-Current (CC) interactions are included, as they distinguish neutrino flavors through coupling with their corresponding charged leptons via $W^{\pm}$ exchange. Neutral-Current (NC) interactions, mediated by the $Z$ boson, are flavor-independent and contribute an equal potential to all flavors, introducing only an overall phase that does not affect oscillation probabilities. Hence, the NC term is omitted  in the effective Hamiltonian without loss of generality.
The simplified version of Eq.(\ref{eq:2.1}) can be written as
\begin{equation}
   \mathcal{H}= \frac{\Delta m_{31}^2}{2E}R_{23}\hs U_{\delta}\hs M \hs U_{\delta}^{\dagger}\hs R_{23}^{T},
\end{equation}
where, $U_{\delta}=\mbox{diag}(1,1,e^{i\delta_{CP}})$ and $M=[R_{13} \hs R_{12} \hs \mbox{diag}(0,\alpha,1)\hs R_{12}^{T}\hs R_{13}^{T}+\mbox{diag}(\Hat{A},0,0)]$.
Diagonalization of this matrix yields the eigenvalues  $E_{1}$, $E_{2}$ and $E_{3}$ and the corresponding eigenvectors $v_{1}$, $v_{2}$ and $v_{3}$. By stacking these eigenvectors, the resulting mixing matrix is given by 
$W=(v_{1}\quad v_{2}\quad v_{3})$. The modified mixing matrix in matter is then expressed as 
$\mathbb{U}=R_{23}\hs U_{\delta}\hs W$. The time-evolved flavor state is given by \cite{Konwar:2024nrd}

\begin{eqnarray}
    \ket{\nu_{\alpha}(t)} &=& \sum_{\beta=e,\mu,\tau}\,\widetilde{\mathbb{U}}_{\alpha \beta}  |\nu_{\beta} \rangle
    \nonumber \\
    &=& \mathbb{U}_{e 1}\mathbb{U}^{*}_{\alpha 1}e^{-iE_{1}t} \ket{\nu_{e}}
    +\mathbb{U}_{\mu 2}\mathbb{U}^{*}_{\alpha 2}e^{-iE_{2}t}\ket{\nu_{\mu}}
    +\mathbb{U}_{\tau 3}\mathbb{U}^{*}_{\alpha 3}e^{-iE_{3}t} \ket{\nu_{\tau}},
\end{eqnarray}

where, $\mathbb{U}_{\alpha i}$'s 
are the component of the matter-modified mixing matrix $\mathbb{U}$ and $\Tilde{\mathbb{U}}_{\alpha\beta}=\mathbb{U}_{\beta i}\mathbb{U}_{\alpha i}^{*}e^{-i E_{i}t}$. This leads to the construction of the time-evolved flavor state taking $\nu_{\mu}$ as initial flavor state,
\begin{equation}
    \ket{\nu_{\mu}(t)}=\Tilde{\mathbb{U}}_{\mu e}\ket{\nu_e}+\Tilde{\mathbb{U}}_{\mu\mu}\ket{\nu_\mu}+\Tilde{\mathbb{U}}_{\mu \tau}\ket{\nu_\tau}.
\end{equation}

 In contrary with the standard concept of entanglement between the two particles, the concept of single particle entanglement (SPE)  has already been introduced in the literature extensively \cite{PhysRevA.72.064306}. 
 Moreover the experimental verification along with the non-local correlation of SPE has also been established by \textit{Pasini et. al.} \cite{PhysRevA.102.063708} in photon system. The core concept of such entanglement was manifested between two different degrees of freedom of a single particle state i.e. spin and orbital angular momentum \cite{science.aat9042} of single particle state. 
 In neutrino system, more precisely, three flavor neutrino system, SPE concept was initiated using single particle multi-mode entanglement between different flavor states, in terms of occupation numbers: $ \ket{\nu_{e}} \equiv \ket{1}_{e}\otimes\ket{0}_{\mu}\otimes\ket{0}_{\tau}$, $ \ket{\nu_{\mu}} \equiv \ket{0}_{e}\otimes\ket{1}_{\mu}\otimes\ket{0}_{\tau}$ and $ \ket{\nu_{\tau}} \equiv \ket{0}_{e}\otimes\ket{0}_{\mu}\otimes\ket{1}_{\tau}$ \cite{Konwar:2024nrd,Li:2021epj,Blasone:2014cub,KumarJha:2020pke} where $\ket{0}_{\nu_{\alpha}}$ and $\ket{1}_{\nu_{\alpha}}$ correspond respectively, to the absence and presence of a neutrino in mode $\alpha$.

 This perspective motivates a further investigation into neutrino entanglement, particularly in the context of the DUNE experiment. For DUNE set up, one prepares muon neutrino beam $\nu_\mu$ which propagates through the medium and finally, evolves into different flavor modes as $\nu_e, \nu_\mu, \nu_\tau$. In this scenario, the density matrix for the state
$\rho_{\mu}(t)=\ket{\nu_{\mu}(t)}\bra{\nu_{\mu}(t)}$, is given by 
\cite{Konwar:2024nrd,Li:2021epj},
\begin{equation}
   \rho_{\mu}(t) =\begin{pmatrix}
          0 & 0 & 0 & 0 & 0 & 0 & 0 & 0\\
          0 & ~|\widetilde{\mathbb{U}}_{\mu\tau}|^{2} ~ & ~\widetilde{\mathbb{U}}_{\mu\tau}\widetilde{\mathbb{U}}^{*}_{\mu\mu} ~&~ 0 ~&~ \widetilde{\mathbb{U}}_{\mu\tau}\widetilde{\mathbb{U}}^{*}_{\mu e} ~&~ 0 ~&~ 0 ~& 0\\
          0 &~ \widetilde{\mathbb{U}}_{\mu\mu }\widetilde{\mathbb{U}}^{*}_{\mu\tau} ~&~ |\widetilde{\mathbb{U}}_{\mu\mu}|^{2} ~&~ 0 ~&~ \widetilde{\mathbb{U}}_{\mu\mu}\widetilde{\mathbb{U}}^{*}_{\mu e} ~&~ 0 & 0 & 0\\
          0 & 0 & 0 & 0 & 0 & 0 & 0 & 0\\
          0 & \widetilde{\mathbb{U}}_{\mu e}\widetilde{\mathbb{U}}^{*}_{\mu\tau} & \widetilde{\mathbb{U}}_{\mu e}\widetilde{\mathbb{U}}^{*}_{\mu\mu} & 0 & |\widetilde{\mathbb{U}}_{\mu e}|^{2} & 0 & 0 & 0\\
          0 & 0 & 0 & 0 & 0 & 0 & 0 & 0\\
          0 & 0 & 0 & 0 & 0 & 0 & 0 & 0\\
          0 & 0 & 0 & 0 & 0 & 0 & 0 & 0\\
      \end{pmatrix}.\\
\label{eq:2.4}
\end{equation}
Clearly from the above expression $|\widetilde{\mathbb{U}}_{\mu\mu}|^{2}=P_{\mu\mu}$, $|\widetilde{\mathbb{U}}_{\mu e}|^{2}=P_{\mu e}$ and $|\widetilde{\mathbb{U}}_{\mu\tau}|^{2}=P_{\mu\tau}$ with the normalization condition $|\widetilde{\mathbb{U}}_{\mu\mu}|^{2}+|\widetilde{\mathbb{U}}_{\mu e}|^{2}+|\widetilde{\mathbb{U}}_{\mu\tau}|^{2}=1$.
The classification of entanglement in a three-qubit system can be categorized as follows:
\begin{enumerate}
    \item Fully separable (no entanglement)
    \item Bipartite entanglement (general bi-separability)
    \item Fully tripartite entanglement (inseparable state or genuine tripartite entanglement)
\end{enumerate}
These classifications are applicable to both pure and mixed states. For a three-partite system, a pure state $\ket{\psi}$ is considered fully separable if it can be expressed as $\ket{\psi}=\ket{\phi}_{A}\otimes\ket{\phi}_{B}\otimes\ket{\phi}_{C}$ \cite{PhysRevA.72.022333}. If the state is not fully separable but can be written as $\ket{\psi}=\ket{\phi}_{A}\otimes\ket{\phi}_{BC}$, it is bi-separable. A state is fully inseparable when it cannot be decomposed in this manner. Bi-separable states, which involve entanglement between a single pair of qubits, exhibit partial bipartite entanglement \cite{PhysRevA.98.052351, Mendes-Santos_2020}. 
The entanglement of reduced two-party states will be then referred to as the reduced entanglement of the pair. This reduced entanglement quantifies the residual two-qubit entanglement when the third qubit is traced out.

The probability analysis suggests that neutrino oscillation in a three-flavor system can be modeled as a three-qubit state. Two relevant formations out of many possibilities in the context of neutrinos are:
\begin{enumerate}
    \item \textbf{Simply separable three-qubit state:} When the separable qubit is traced out, the remaining two qubits remain entangled; $\ket{\psi}=\ket{\phi}_{A}\otimes\ket{\phi}_{BC}$ where $\ket{\phi}_{BC}$ entangled, results in an entangled reduced state $\rho^{BC}=\ket{\phi}_{BC}\bra{\phi}_{BC}$. 
    This type of state also exists in non-pure forms. Such states, referred to as W states, exhibit robust entanglement that persists even when one of the three qubits is lost.
    \item \textbf{Fully inseparable state:} In this scenario, all three reduced entanglements are non-zero. Such states, referred to as GHZ states.
\end{enumerate}

 Neutrino Oscillation is an intrinsic quantum mechanical phenomenon that deals with the three objects (three flavored states of neutrino) which belong to the equal dimensional but different Hilbert spaces $\mathcal{H}_{1}, \mathcal{H}_{2}$ and $\mathcal{H}_{3}$. However, the density matrix of the combined system lies in the Hilbert space $\mathcal{H}_{1}\otimes\mathcal{H}_{2}\otimes\mathcal{H}_{3}$ with dimensionality 8. For a bipartite system, we say that $\rho_{12}$ (the density matrix for the bipartite system) is not entangled i.e., separable, if it is possible to write $\rho_{12}=\sum_{j}\lambda_{j}\hspace*{0.03 true in} \rho_{1}^{j}\otimes \rho_{2}^{j}$\hspace*{0.01 true in}, ~$\forall ~\lambda_{j}>0$. If such decomposition is impossible, we say that the system is entangled. However, analyzing entanglement in a tripartite system using this decomposition method is challenging. We can use the positive partial transpose (PPT) criterion to determine the presence of entanglement in a three-party system.
 In the partial transpose density matrix, if at least one negative eigenvalue is found, it indicates the system is entangled. The PPT criterion is one approach to assess this \cite{Plenio:2007zz, RevModPhys.81.865}. 
This raises a further question about the measure of the entanglement as well as its nature: whether it is genuine tripartite or partial bipartite entanglement. While extensive research has been done on this topic, we will focus on practical justifications for the experimental setup of DUNE. We introduce two major entanglement measures: EOF and concurrence to quantify the entanglement.

\textit{\textbf{Entanglement of formation (EOF):}}
The entanglement of formation is the most basic measurement of entanglement. Given a density matrix $\rho$ of a pair of quantum systems A and B, consider all possible pure state decompositions of $\rho$, i.e., all ensembles of states $\ket{\psi_{i}}$ with probabilities $p_{i}$, such that
\begin{equation}
    \rho=\sum p_{i}\ket{\psi_{i}}\bra{\psi_{i}}.
\label{eq:2.6}
\end{equation}
For each pure state, the entanglement of formation is defined as the entropy of either of the two systems A and B,
\begin{equation}
    {\rm EOF}(\psi)=-\mbox{Tr}(\rho_{A}\log_{2}\rho_{A})=-\mbox{Tr}(\rho_{B}\log_{2}\rho_{B}).
\label{eq:2.6}
\end{equation}
Here, $\rho_{A}$ is the partial trace of $\ket{\psi}\bra{\psi}$ over subsystem B, and $\rho_{B}$ is defined similarly \cite{PhysRevLett.70.1895, PhysRevA.54.3824}. From (\ref{eq:2.6}), the term $\mbox{Tr}(\rho_{A}\log_{2}\rho_{A})$ effectively suggests the entropy of the pure system $A$. Consequently, the expression for the relative entropy for a bipartite system is $-\sum_{i}k_{i}^{2}\log_{2}k_{i}^{2}$ \cite{PhysRevLett.80.2245}, where $k_{i}$ is the Schmidt coefficient and satisfies the condition $\sum_{i}k_{i}^{2}=1$.
Analogically, with the neutrino system the Schmidt coefficient $k^{2}_{i}$ leads to the appearance and disappearance probabilities $P_{a}$ and $P_{d}$ adhering the property ($P_{d}+\sum P_{a})=1$.
This observation leads to the same conclusion about the conservation of probabilities.
For a tripartite system, the EOF is given by \cite{PhysRevA.99.042305, PhysRevA.101.032301},
\begin{equation}
\begin{split}
    {\rm EOF} &=(S_{A}+S_{B}+S_{C}),\\
\end{split}
\end{equation}
where the individual component $S_{A},S_{B}$ and $S_{C}$ gives the entropy associated with the different flavor states in three flavor neutrino oscillations with the expression, 
$S_{A}=-\frac{1}{2}\big[P_{\alpha \beta}
\\ \log_{2}P_{\alpha\beta}
+(P_{\alpha\alpha}+P_{\alpha\gamma})\log_{2}(P_{\alpha\alpha}+P_{\alpha\gamma})\big]$, 
    $S_{B}=-\frac{1}{2}\big[P_{\alpha\alpha}\log_{2}P_{\alpha\alpha}+(P_{\alpha\beta}+P_{\alpha\gamma})\log_{2}(P_{\alpha\beta}+P_{\alpha\gamma})\big]$ and 
  $S_{C}=-\frac{1}{2}\big[P_{\alpha\gamma}\log_{2}P_{\alpha\gamma}+(P_{\alpha\alpha}+P_{\alpha\beta})\log_{2}(P_{\alpha\alpha}+P_{\alpha\beta})\big]$. Here, $\alpha=e,\mu,\tau$ corresponds to the initial flavor of the neutrino and $\beta, \gamma = e, \mu, \tau$ represent the other two oscillating flavors. For the present work, we consider the $\nu_{\mu}$ as the initial flavor for the typical production mechanism of accelerator neutrinos.

\textit{\textbf{Concurrence:}} Concurrence is a quantitative measure of entanglement for a bipartite system, defined as \cite{PhysRevLett.80.2245}:
\begin{equation}
    C^{\alpha}=\Big[\mbox{max}(\lambda_{1}-\lambda_{2}-\lambda_{3}-\lambda_{4},0)\Big],
\end{equation}
where $\lambda_{i}$ are the eigenvalues of the reduced density matrix of the bipartite system. In the context of two-flavor neutrino oscillations, the concurrence can be expressed as $C_{\alpha\beta}=2\sqrt{\mbox{P}_{\alpha \beta}\mbox{P}_{\alpha \alpha}}$, $\mbox{P}_{\alpha \beta}$ and $\mbox{P}_{\alpha \alpha}$ \cite{KumarJha:2020pke} represent the appearance and disappearance probabilities, respectively. Recent studies \cite{Blasone:2007vw,Blasone:2010ta, Blasone:2013zaa, Blasone:2014jea,Blasone:2015lya} have demonstrated that the concept of concurrence can be extended to quantify entanglement in three-qubit states as well, significantly advancing our understanding of entanglement in more complex quantum systems.
Although the expression and analysis of the concurrence of the bipartite system have already been discussed vividly in the literature, the expression for the concurrence of the tripartite system was proposed by Guo \textit{et al.} \cite{PhysRevA.99.042305}
\begin{equation}
    C^{\alpha}=\big[3-{\rm Tr}(\rho_{\alpha})^{2}-{\rm Tr}(\rho_{\beta})^{2}-{\rm Tr}(\rho_{\gamma})^{2}\big]^{1/2}.
\label{eq:2.10}
\end{equation}
For the three-flavor neutrino system Eq. (\ref{eq:2.10}) modifies to the expression,
\begin{equation}
    C^{\alpha^{2}}=C^{2}_{\beta\alpha;\alpha}+C^{2}_{\gamma\alpha;\alpha}+4\mbox{P}_{\alpha\beta}\mbox{P}_{\alpha\gamma}\;.
\end{equation}
Here, $\alpha$ denotes the initial flavor of the neutrino, while $\beta$ and $\gamma$ represent the oscillated flavor states. The maximum value of $C^{\alpha^2}$ corresponds to the maximal entanglement of the system, whereas the minimum value indicates the separability of the states.  $C^2_{\beta (\gamma) \alpha;\alpha}$ denotes the measurement of entanglement between the initial flavor state $\nu_{\alpha}$ and oscillating flavor state $\nu_{\beta}$ ($\nu_{\gamma}$). The last term $\mbox{P}_{\alpha\beta (\gamma)}$ represents the oscillation probability of $\nu_{\alpha}$ to $\nu_{\beta} \hs
 (\nu_\gamma)$.

 \textit{\textbf{Entanglement Monogamy:}}
 Using squared concurrence (SC) as an entanglement measure, Coffman, Kundu, and Wootters established the first monogamy inequality for three-qubit states, now known as the CKW inequality. Consider a three-particle system ABC. If we treat the pair BC as a single composite system, we can define the concurrence $C_{A|BC}$ between qubit A and the pair BC. For this purpose, A and BC can be treated as a pair of qubits in a pure state. The CKW inequality is then expressed as:
\begin{equation}
    C^{2}_{AB}+C^{2}_{AC}\leq C^{2}_{A|(BC)}\;.
\label{eq:2.12}
\end{equation}
$C_{AB}^{2}$ and $C^{2}_{AC}$ are defined as the squared entanglement between $(A;B)$ and $(A;C)$ respectively \cite{Coffman_2000}.
 Informally, the above equation can be interpreted as follows: Qubit A shares a certain amount of entanglement with the combined system BC. This total entanglement places an upper limit on how much entanglement A can share with qubits B and C individually. Furthermore, the portion of entanglement that A shares with qubit B is not available for sharing with qubit C, highlighting the restricted shareability of entanglement in quantum systems.
 Notably, the entanglement of formation, in its unsquared form, does not satisfy the monogamy relation. However, similar monogamy inequalities have been successfully derived for other squared measures, such as the squared entanglement of formation (SEF).
Unlike classical correlations, quantum entanglement cannot be freely shared among multiple parties. This limitation, known as the monogamy of entanglement, is one of the most fundamental properties of quantum entanglement. A key distinction between quantum entanglement and classical correlation is this restricted shareability: in a multi-party quantum system, if two parties are maximally entangled, neither can share entanglement with any other part of the system. This inherent constraint on the distribution of entanglement is what defines the monogamy of entanglement.

A fundamental question in the study of the MOE is determining whether a given entanglement measurement exhibits the monogamous property. There are various approaches to defining this property for an entanglement measure. Typically, a monogamy relation for an entanglement measurement $E$ is expressed quantitatively,
\begin{equation}
    E(\rho_{A|BC})\geq E(\rho_{AB})+E(\rho_{AC}),
\end{equation}
where $E(\rho_{A|BC})$  is an entanglement measure that quantifies the degree of entanglement between A and the combined system BC. This inequality implies that the total entanglement between A and each of the other parties, B or C, cannot exceed the entanglement between A and BC.
\\ In the context of tripartite systems, there are three distinct types of separability:
\begin{itemize}
\item Fully separable state
\item Two-partite separable state
\item Genuinely entangled state
\end{itemize}
 Informally, the equation can be interpreted as follows: Qubit A shares a certain amount of entanglement with the combined system BC. This total entanglement places an upper limit on how much entanglement A can share with qubits B and C individually. Furthermore, the portion of entanglement that A shares with qubit B is not available for sharing with qubit C, highlighting the restricted shareability of entanglement in quantum systems.
MOE is a fundamental characteristic of quantum entanglement. 
 The restriction from MOE highlights one of the key differences between quantum entanglement and classical correlation \cite{Henderson:2001wrr}.

MOE limits the amount of information that an eavesdropper can potentially extract during secret key extraction, thereby playing a crucial role in ensuring the security of quantum key distribution. Additionally, MOE has significant applications across various domains of physics, including the classification of quantum states \cite{Zong:2022xza}, no-signaling theories \cite{Ryu_2021}, condensed-matter physics \cite{Osterloh_2015}, and even black-hole physics \cite{PhysRevD.107.066020}.
\section{Experimental setup and simulation details}
\label{exp}
Deep Underground Neutrino Experiment (DUNE) is one of the most promising upcoming long-baseline neutrino experiment. It has a baseline of 1300 km from the neutrino source at Fermi National Accelerator Laboratory (FNAL) to Sanford Underground Research Facility (SURF). For simulating the DUNE experiment, we use the files provided in the Technical Design Report paper \cite{DUNE:2021cuw}. The files correspond to $624$ kt-MW-years of exposure; 6.5 years each in neutrino  and antineutrino  modes, with a 40 kt fiducial mass liquid argon time-projection chamber (LArTPC) far detector and a 120-GeV, 1.2 MW beam. This setup is equivalent to ten years of data collection based on the nominal staging assumptions described in \cite{DUNE:2020jqi}.

We have used General Long-Baseline Experiment Simulator (GLoBES) \cite{Huber:2004ka, Huber:2007ji} software package to simulate the DUNE experiment. To calculate the sensitivities, we applied  Poisson log-likelihood formula, given by: 
\begin{equation}
    \chi^2 = 2 \sum_{i=1}^n \left[N_i^{\rm test} - N_i^{\rm true} - N_i^{\rm true} \rm{log} \left( \frac{N_i^{ \rm test}}{N_i^{\rm true}} \right) \right],
    \label{chi}
\end{equation}
where, $N_{i}^{\rm true}$ and $N_{i}^{\rm test}$ represent the true and test sets of event numbers, respectively, and `$i$' denotes the number of energy bins. For sensitivity calculations, we used the true values of the oscillation parameters listed in Table \ref{osc11}. We minimized over all six oscillation parameters to obtain the minimum $\chi^2$ value.

\begin{table}[]
    \centering
    \begin{tabular}{c c c c c}
    \hline
    \hline
      $\theta_{12}$   & $\theta_{13}$ & $\theta_{23}$ & $\Delta m_{21}^2  (\rm eV^2)$ & $\Delta m_{31}^2 (\rm eV^2)$ \\
      $33.68^{\circ}$ & $8.56^{\circ}$ &  $43.3^{\circ}$ & $7.49 \times 10^{-5}$ & $\pm 2.513 \times 10^{-3}$ \\
        \hline
        \hline
    \end{tabular}
    \caption{Oscillation parameters used to plot the probability and EOF, concurrence measurement \cite{Esteban:2024eli}}.
    \label{osc11}
\end{table}

\section{Analysis}
\label{analysis}
\subsection{Verification of entangled state and quantification}
\subsubsection{Entanglement of formation}
\begin{figure}[htb!]
\vspace{0.5cm}
\includegraphics[scale=0.20]{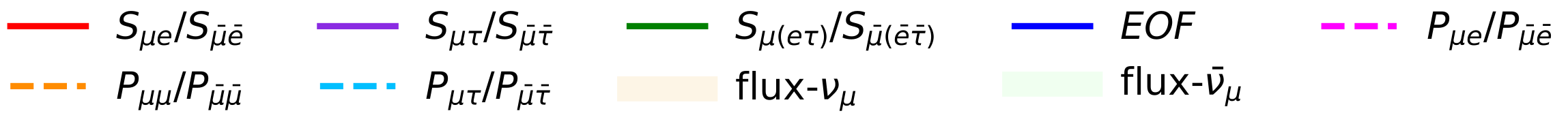}\\
\includegraphics[scale=0.5]{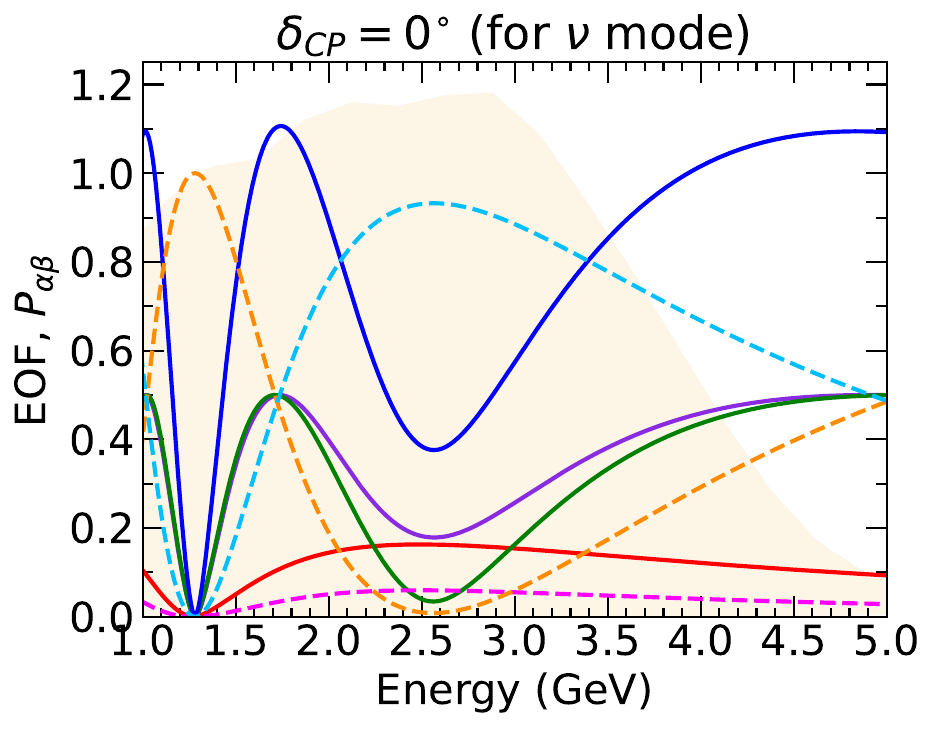}	\includegraphics[scale=0.5]{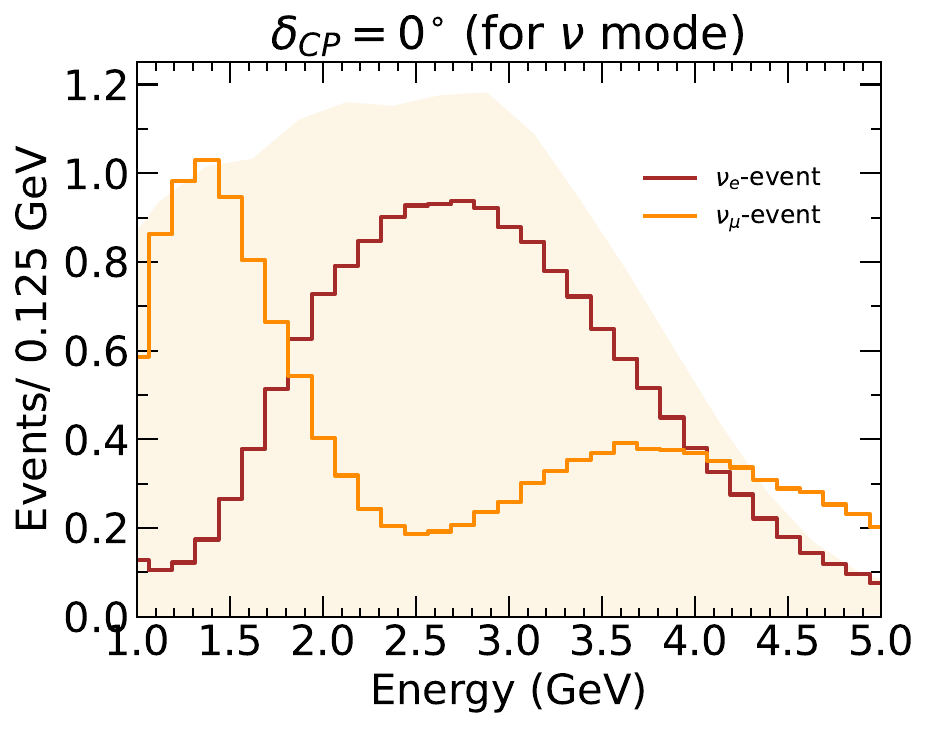}\\
\includegraphics[scale=0.5]{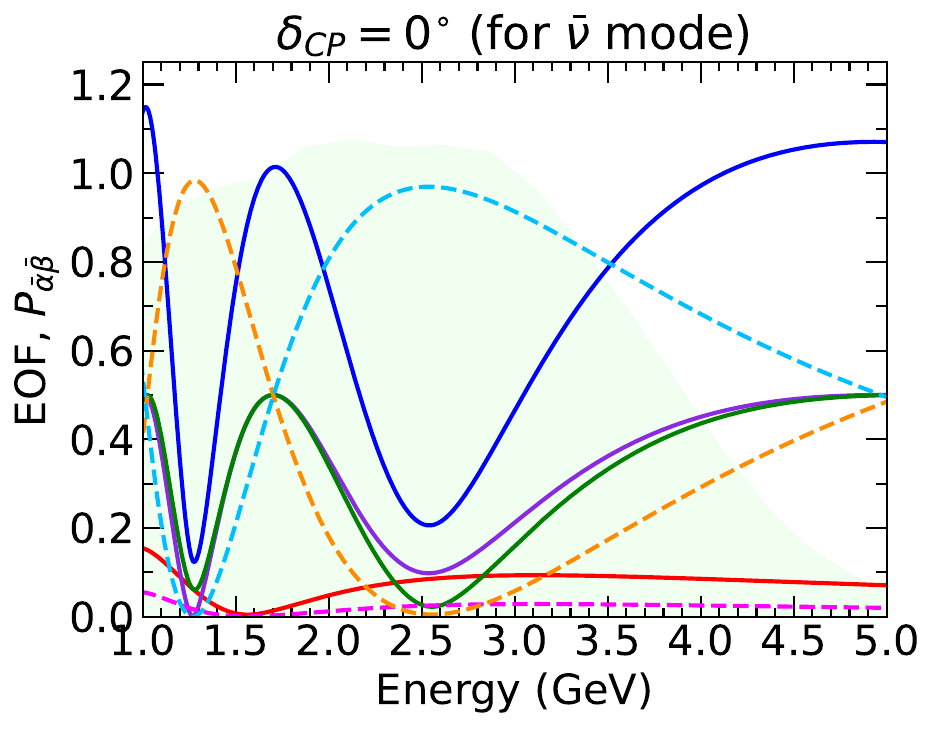}
\includegraphics[scale=0.5]{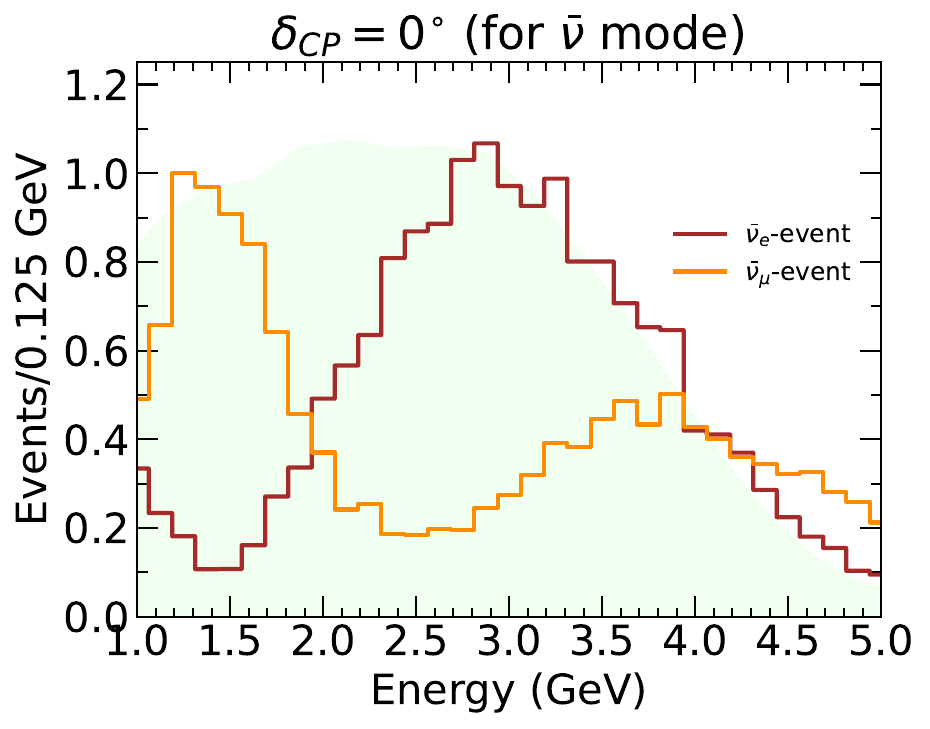}
\caption{Left column depicts the EOF measurements and probability curves with respect to neutrino energy for DUNE experimental configuration. Right column shows the appearance and disappearance  event rates. Color codes are given in the legend. Upper (lower) row is for (anti) neutrino mode. }
	\label{Fig.1}
\end{figure}

In the first hand we have used the entanglement of formation (EOF) as entanglement measures because from the mathematical expression of EOF we can see that it is directly related to the linear entropy of the individual components of the system which is the fundamental measure of a mixed system \cite{PhysRevA.67.022110}. Hence using EOF as a entanglement measures justifiably signifies the mixedness within the neutrino system and the amount of entanglement present in it.
In the experimental setup of DUNE, as described in the previous section, we first verify entanglement in the framework of EOF, the theoretical aspects of which has been discussed in Section \ref{theory}. The accelerator-based neutrino experiment predominantly generates muon-type neutrinos, establishing the initial flavor state as $\nu_{\mu}$ for this study. The goal is to quantify the entanglement between $\nu_{\mu}$ and the other two flavor states, $\nu_{e}$ and $\nu_{\tau}$. For the initial flavor state $\nu_{\mu}$, the EOF expression simplifies to,
\begin{equation}
   {\rm  EOF}= S_{\mu e}+S_{\mu\tau}+S_{\mu(e\tau)}.
\label{eq:4.1}
\end{equation}
Here, the first term,  $S_{\mu e}$, represents the entanglement between   $\nu_{\mu}$ and $\nu_{e}$, while the second term, $ S_{\mu \tau} $, describes the entanglement between $\nu_{\mu}$ and $\nu_{\tau}$. The final term, $ S_{\mu(e\tau)} $, accounts for the entanglement of the biseparable state $\nu_{\mu}-\nu_{(e,\tau)}$. 
The explicit expressions for $ S_{\mu e} ,  S_{\mu \tau} $, and $ S_{\mu(e\tau)} $ are given by:
\begin{equation}
\begin{split}
    &S_{\mu e} =-\frac{1}{2} \Big[P_{\mu e}\log_{2}P_{\mu e}+(P_{\mu\mu}+P_{\mu\tau})\log_{2}(P_{\mu\mu}+P_{\mu\tau})\Big],\\
    &S_{\mu \tau} =- \frac{1}{2} \Big[ P_{\mu\tau}\log_{2}P_{\mu\tau}+(P_{\mu\mu}+P_{\mu e})\log_{2}(P_{\mu\mu}+P_{\mu e}) \Big],\\
    &S_{\mu (e\tau)} =- \frac{1}{2} \Big[ P_{\mu\mu}\log_{2}P_{\mu\mu}+(P_{\mu e}+P_{\mu\tau})\log_{2}(P_{\mu e}+P_{\mu\tau})\Big].\\
\end{split}
\label{eq.4.2}
\end{equation}
Fig. \ref{Fig.1} illustrates the EOF measurement in the DUNE experimental configuration. Each plot in Fig. \ref{Fig.1} is generated using the oscillation parameter values provided in Table \ref{osc11}, with the CP-violating phase $\delta_{CP}$, set to $0^{\circ}$.  
The left column displays the EOF measurements, correlating with both appearance and disappearance probabilities as a function of neutrino energy. The upper row of Fig. \ref{Fig.1} depicts results for neutrino mode, while the lower row shows results for anti-neutrino mode. In the left column, each panel contains seven curves: the red, purple and green curves correspond to the three terms, $ S_{\mu e} ,  S_{\mu \tau} $, and $ S_{\mu(e\tau)} $  from Eq. \ref{eq:4.1}. The blue solid line represents the sum of these terms, i.e. the total EOF of the three flavor neutrino system. Additionally, the pink dashed curve indicates the appearance probability for electron neutrinos, the orange dashed curve represents the disappearance probability for muon neutrinos, and the cyan dashed curve shows the appearance probability for tau neutrinos.

From the panels, we can observe three main energy windows where the EOF is achieving two minima and one maximum. However, one of those two minima  tends to zero and we define this minimum as global minimum. For the other minimum, we designate it as local minimum. The EOF global minimum is seen in the energy range  ($1.0$ - $1.5$) GeV, the maximum from $1.5$ GeV to $2.0$ GeV and the local minimum from $2.0$ GeV to $3.0$ GeV energy ranges. For our convenience, we describe these three energy windows as:    
\begin{itemize}
\item Energy window (EW1): $1.0-1.5$ GeV
\item Energy window (EW2): $1.5-2.0$ GeV
\item Energy window (EW3): $2.0-3.0$ GeV
\end{itemize}
Now, let's illustrate each energy window in detail.\\

\textbf{EW1:} 
In the energy range of $1.0$ GeV to $1.5$ GeV, the EOF reaches a minimum, approaching nearly zero for both neutrino and anti-neutrino modes. A zero EOF value confirms the complete separability of one flavor state from the others. To identify the separable flavor state, we examine the appearance and disappearance probabilities alongside the EOF curves. In this energy range, $\nu_{\mu}$ disappearance reaches its maximum, while $P_{\mu e}$ and $P_{\mu \tau}$ are zero, indicating that the $\nu_{\mu}$ flavor state is nearly entirely separable from the other two flavors. Notably, at $1.27$ GeV, the EOF value is exactly (or nearly) zero for neutrino (anti) modes, which we refer to as the ``global minimum" point throughout our analysis.
This global minimum point offers an advantage for the precise measurement of neutrino oscillation parameters, which we will explore in the following section. Additionally, within the EW1 range, all three components of the EOF measurement (given in Eq. \ref{eq.4.2}) reach their minimum, further indicating a pure state configuration in this energy window.

 \textbf{EW2}: The energy window from $1.5$ GeV to $2.0$ GeV shows a peak in the total EOF value. The entanglement measure reaches its maximum when the appearance and disappearance probabilities of the bipartite system are equal, causing the system to behave as a maximally mixed state. Within this range, around $1.74$ GeV, we observe that both $P_{\mu\tau}$ and $P_{\mu\mu}$ equal to 0.5. At this point, $S_{\mu\tau}$ and $S_{\mu(e \tau)}$ reach their maximum values, indicating the emergence of maximal entanglement, particularly between $\nu_{\mu}$ and $\nu_{\tau}$. 
The upper bound of the EOF is given by $\log_2 d$, where $d$ denotes the dimension of the Hilbert space of each subsystem \cite{Bennett:1996gf}. For a two-qubit system $(d = 2)$, the maximum EOF is $\log_2 2 = 1$. In contrast, for neutrinos, where $d = 3$, the upper limit becomes $\log_2 3 \approx 1.58$, which exceeds 1. Hence, obtaining EOF values greater than 1 in this case is physically consistent.  
In this energy range, all three flavor states are in maximally entangled state with a tendency of maximal mixing between $\nu_{\mu}$ and $\nu_{\tau}$, making it difficult to extract ``pure" information about the oscillation parameters.

 \textbf{EW3}:
At an energy of 2.5 GeV, we observe minima in both $S_{\mu\tau}$ and $S_{\mu(e\tau)}$. This can be physically interpreted as the entanglement measure reaching its lowest point when the appearance probability is at its maximum. This suggests that either the two-flavor states are nearly separable or the entanglement is not at its maximum. 
Within the $2.0-3.0$ GeV energy window, these minima indicate that the states are neither fully separable nor maximally entangled. The non-zero minima imply that, while the entanglement is not maximal, there is still a degree of entanglement present. The non-zero value confirms that some level of entanglement persists within the system. 
Additionally, at approximately $E_{\nu} = 2.5$ GeV, the total EOF of the system also reaches a minimum. We interpret this minimum as a ``local minimum" due to the non-zero entanglement value, which indicates that the system retains a degree of entanglement even at this energy.  The lower panel of Figure \ref{Fig.1} displays corresponding results for antineutrinos, which exhibit similar trends to those seen for neutrinos. 

Although, the EOF provides an interesting aspects in the entanglement measures in a mixed quantum system, it is very hard to compute. As EOF is a logarithmic function, it is very difficult to analyze the nature of the entanglement (bipartite or pure tripartite entanglement) within the system.
 It is noteworthy that $S_{\mu e}$ reaches a maximum at 2.5 GeV, which is counterintuitive given the expected trend for an entangled flavor state which is hard to explain.
 
 The above analysis suggests clearly that although it depicts a hint of maximal entanglement between muon and tau flavor state of neutrino but due to logarithmic nature of EOF, we can not conclusively say which flavor states are entangled. Hence we proceed further to analyse with respect to examine the nature of  entanglement in terms of squared-concurrence.

\subsubsection{Concurrence}
The concurrence for three-flavor neutrino oscillations is defined by the following expression:
\begin{equation}
    C^{\mu^{2}}=C_{\mu e; \mu}^{2}+C_{\mu\tau;\mu}^{2}+4\mbox{P}_{\mu e}\mbox{P}_{\mu\tau},
    \label{con}
\end{equation}
where the term $C_{\mu e; \mu}^{2}$ represents the concurrence between $\nu_{\mu}$ and $\nu_{e}$, and is given by $4P_{\mu e}P_{\mu\mu}$, the term $C_{\mu\tau;\mu}^{2}$ represents the concurrence between $\nu_{\mu}$ and $\nu_{\tau}$,  expressed as $4P_{\mu\tau}P_{\mu\mu}$ while the third term $4 \text{P}_{\mu e}\text{P}_{\mu\tau}$ describes the product of two oscillation probabilities, which we redefine as $C^{2}_{\mu e\tau}$.

\begin{figure} 
\vspace{0.5cm}
\includegraphics[scale=0.2]{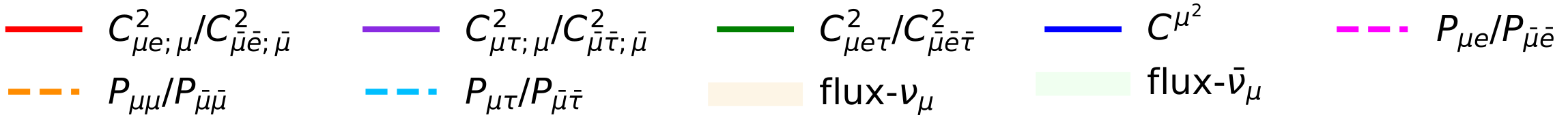}\\
\includegraphics[scale=0.5]{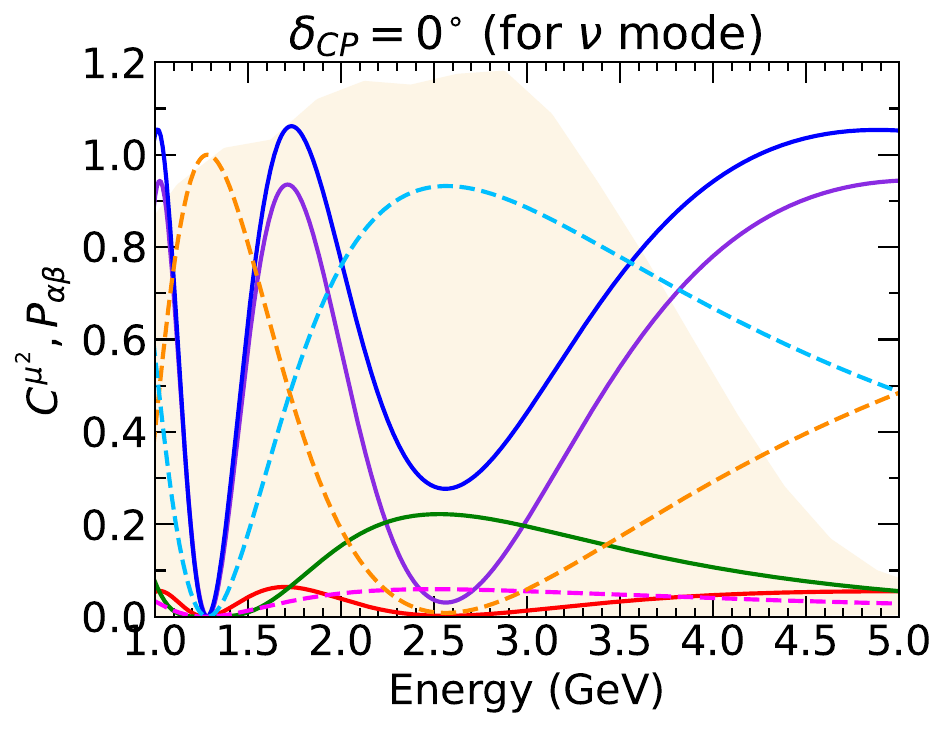}
    \includegraphics[scale=0.5]{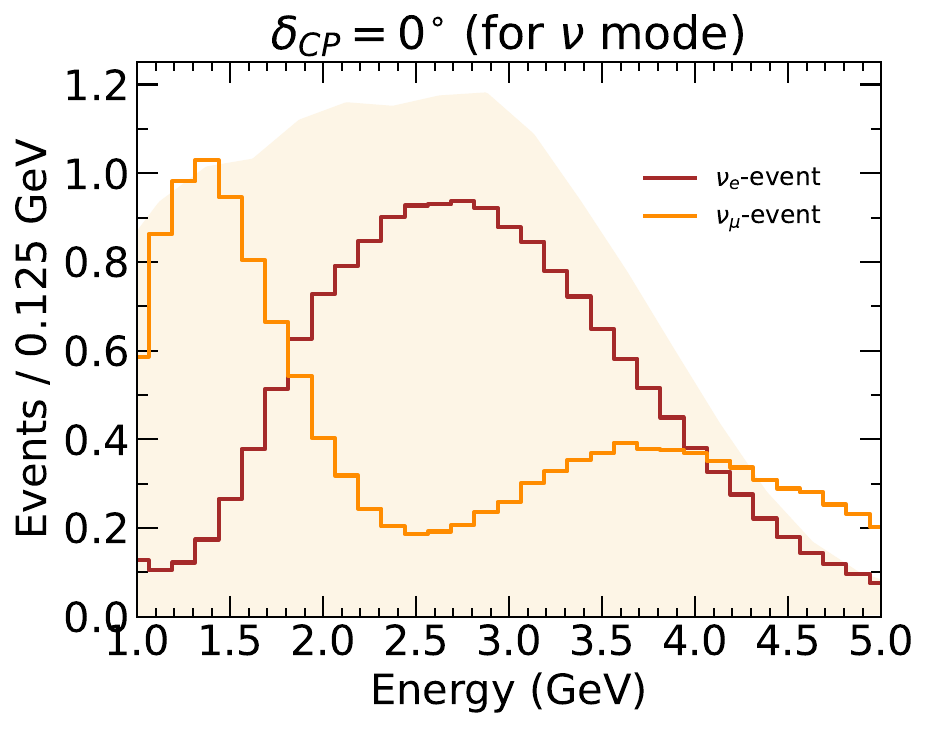}\\
    \includegraphics[scale=0.5]{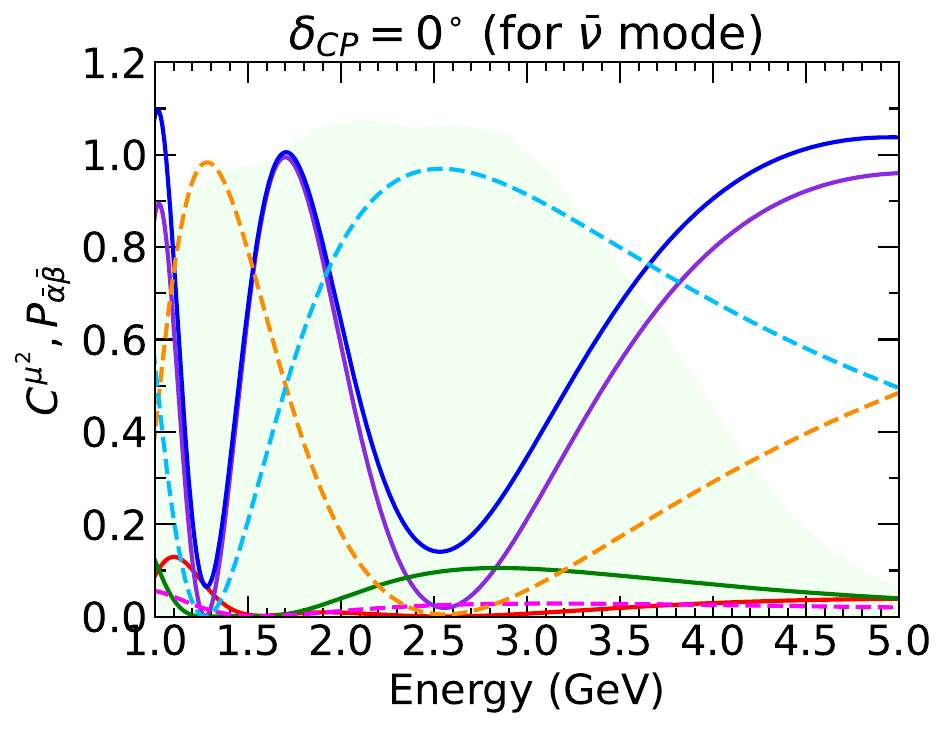}
    \includegraphics[scale=0.5]{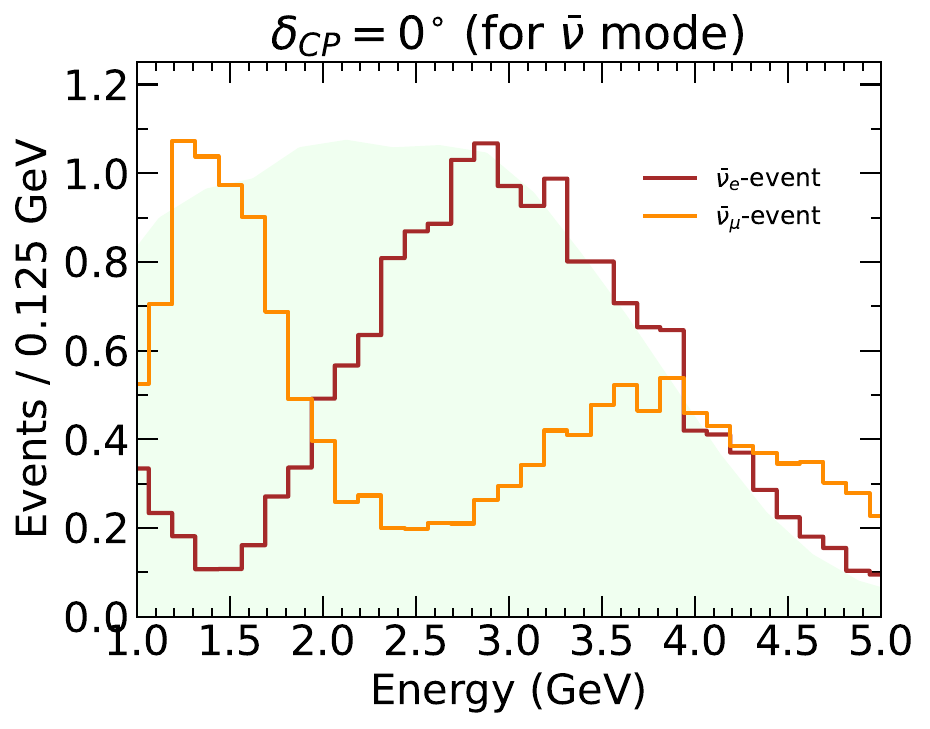}\\
    \hspace*{7.5cm}
    \caption{Left column depicts the concurrence measurements and probability curves with respect to neutrino energy for DUNE experimental configuration. Right column shows the appearance and disappearance event rates. Color codes are given in the legend. Upper (lower) row is for (anti) neutrino mode.}
    \label{fig:2}
\end{figure}

By plotting these three terms individually, we can visualize the contribution of each flavor state to the total concurrence. This visualization aids in understanding the contributions from each flavor state and helps to justify the monogamy inequality, which indicates strong correlations between flavor states and allows us to identify and characterize specific entangled neutrino flavors. 

In Fig. \ref{fig:2}, the left column displays the concurrences as a function of energy along with the corresponding probabilities. In order to generate the plots, we have taken the values of neutrino oscillation parameters from Table \ref{osc11} with $\delta_{CP}$ as $0^{\circ}$. Seven curves are shown: the red, purple and green curves represent the three concurrence terms, $C_{\mu e; \mu}^{2}$, $C^{2}_{\mu\tau;\mu}$ and $C^{2}_{\mu e\tau}$ respectively from Eq. \ref{con}, the blue curve shows the total concurrence, $C^{\mu^{2}}$ for the DUNE experimental setup, and the pink (orange) and cyan dashed curves represent the appearance (disappearance) probabilities. The analysis will focus on the three distinct energy windows discussed previously.\\

\textbf{EW1:}  In the energy window of $1.0–1.5$ GeV, we observe that the appearance probability $P_{\mu\mu}$ is 1, while $P_{\mu e}$ and $P_{\mu\tau}$ approaching to zero. This behavior indicates that the $\nu_{\mu}$ flavor state is completely disentangled from both $\nu_e$ and $\nu_{\tau}$, resulting in the three neutrino flavor states behaving as a completely separable system. This observation underscores the physical relevance of concurrence in our analysis, as it effectively captures the disentangled nature of the system within this energy range. Effectively it is noticeable in this energy window all terms in Eq. \ref{con} approaches to zero which leads to disentangled state.\\

\textbf{EW2:} In the energy window between 1.5 and 2.0 GeV, there is a specific point, around 1.74 GeV, where $P_{\mu\tau}$ and $P_{\mu\mu}$ become equal. At this point, the total squared concurrence associated with tripartite neutrino system, $C^{\mu^2}$, reaches its maximum value, indicating a high degree of entanglement within the system. This maximum entanglement corresponds to a maximally mixed state, as evidenced by $P_{\mu\mu} = P_{\mu\tau} = 0.5$. The plot shows that this maximal mixing is primarily due to the $\nu_{\mu}$ and $\nu_{\tau}$ states. Examining the individual contributions, we find that the entanglement measures $C^{2}_{\mu\tau; \mu}$ and $C^{2}_{\mu e; \mu}$ both reach peak values at a specific point in the energy spectrum. Notably, $C^{2}_{\mu\tau; \mu}$ exhibits a significantly higher amplitude compared to $C^{2}_{\mu e; \mu}$, indicating a stronger entanglement between the $\nu_{\mu}$ and $\nu_{\tau}$ states. It is interesting to note that, the total concurrence value exceeds the value 1. In our analysis, $C^{\mu^2}$ represents the total concurrence, not the concurrence for just a single bi-partition. For each bi-partition, the concurrence is still less than or equal to one. When we add the three concurrences for three different bi-partitions, it can exceed value 1. The upper bound on total $C^{\mu^2}$ should be $ \frac{2 (d-1)}{d}$ with $d$ being the dimension of the Hilbert space of each subsystem \cite{Li:2005mbb,Dur:2000zz}. For instance, in the case of two qubits (d=
2), the maximum possible value of $C^{\mu^2}$ is 1. However, for three-dimensional systems such as neutrinos (d=3), the upper bound increases to $\frac{2 \times (3-1)}{3}= \frac{4}{3}$ which is greater than 1. Consequently, the total concurrence in our results correctly attains this maximum value of $\frac{4}{3}=1.33$.\\

\textbf{EW3:} In the energy window 2.0–3.0 GeV, where we notice a dip in the values of $C^{2}_{\mu\tau; \mu}$, $C^{2}_{\mu e; \mu}$, and $C^{\mu^2}$. This dip suggests that the entanglement is reduced in this range, indicating a shift away from maximal entanglement.

At approximately 2.5 GeV, the probability $P_{\mu\tau}$ reaches its peak, accompanied by a slight but noticeable increase in $P_{\mu e}$ and an almost vanishing value of $P_{\mu\mu}$. Despite this, $P_{\mu\tau}\neq 1$ at this energy, the flavor states are not fully disentangled; however, they do lose their maximally entangled state. This observation underscores that the nature of the entanglement is predominantly governed by the interaction between the dominant flavor states, $\nu_{\mu}$ and $\nu_{\tau}$. Based on our discussion of the concurrence, we conclude that in this energy window, the three-flavor neutrino system behaves as a bipartite state, with $\nu_{e}$ being in a separable state.

Throughout the entire energy range, the term represented by the green line, $4P_{\mu e}P_{\mu\tau}$, does not directly contribute to the physical interpretation of concurrence but influences the overall concurrence value. This term can be expressed as $\frac{C^{2}_{\mu e; \mu}C^{2}_{\mu\tau; \mu}}{4P^{2}_{\mu\mu}}$, representing the product of the squared concurrences of the $\nu_{\mu}-\nu_e$ and $\nu_{\mu}-\nu_\tau$ pairs. Despite its lack of direct physical significance in terms of concurrence, this term is crucial for validating the monogamy inequality of entanglement, which imposes constraints on how entanglement is distributed among multiple flavor states.

\begin{figure}  
\centering
\includegraphics[scale=0.5]{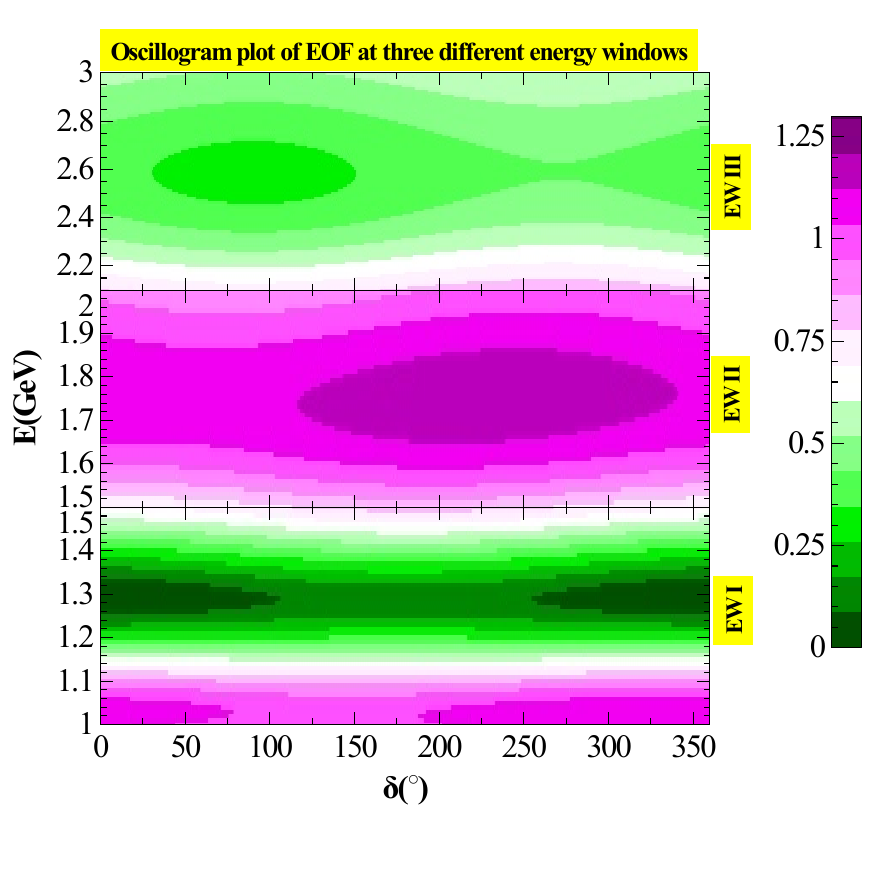}
\includegraphics[scale=0.5]{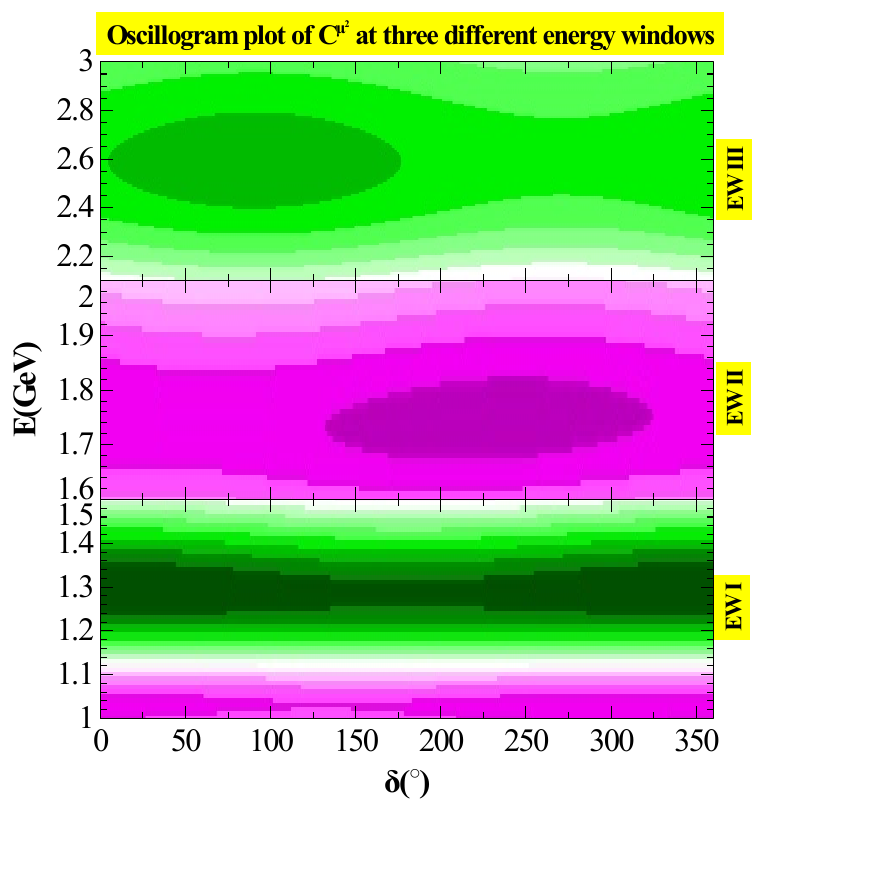}
\vspace{-0.3cm}
    \caption{Left (right) panel shows the oscillogram plot of EOF (concurrence) at three different energy windows for whole $\delta_{CP}$ range.}
    \label{osc}
\end{figure}

In summary, by starting with the initial flavor state $\nu_{\mu}$, our analysis focuses on the bipartite entanglement between $\nu_{e}$ and $\nu_{\tau}$. Our findings demonstrate that, with $\nu_{\mu}$ as the initial flavor state, there is significant entanglement with $\nu_{\tau}$, including instances of maximal entanglement, while $\nu_e$ remains a separable state in certain energy ranges. This highlights the critical role of $\nu_{\mu}$ in the entanglement dynamics of the three-flavor system, particularly in its correlation with $\nu_{\tau}$.

For the analysis of EOF and concurrence, we have used a fixed value of $\delta_{CP} = 0^\circ$. However, the general behavior of the entanglement measure plots as a function of neutrino energy remains consistent for other values of $\delta_{CP}$. To examine the dependency of EOF and concurrence on $\delta_{CP}$, an oscillogram is presented in Figure \ref{osc}. The left panel of the figure illustrates the variation of EOF, while the right panel shows the variation of concurrence for the three previously discussed energy windows across different $\delta_{CP}$ values. In both panels, dark green represents the minimum value of the entanglement measure, while dark pink indicates the maximum values of EOF and concurrence. Both panels reveal that around 1.27 GeV, a global minimum occurs, at 1.74 GeV, a maximum is observed, and at 2.5 GeV, a local minimum appears. This behavior remains consistent across the entire range of $\delta_{CP}$ values. Therefore, although $\delta_{CP} = 0^\circ$ was used in Figures \ref{Fig.1} and \ref{fig:2}, the result remains similar for other values of $\delta_{CP}$ as well.

\subsection{Verification of Entanglement Monogamy}

In Section \ref{theory}, we discussed the mathematical formalism of the CKW inequality. The fundamental expression of this inequality in terms of squared concurrence is given by Eq. \ref{eq:2.12}. For the present analysis, we will investigate CKW inequality for the neutrino system where $A\equiv \nu_{\mu}$, $B\equiv \nu_{e}$ and $C\equiv \nu_{\tau}$. 
It is straightforward to show that $C^{\mu^{2}}\equiv C^{2}_{A|(BC)}$ and we have already shown $C^{2}_{\mu\tau;\mu}=4P_{\mu\tau}P_{\mu\mu}\equiv C^{2}_{AB}$ and $C^{2}_{\mu e;\mu}=4P_{\mu e}P_{\mu \mu}\equiv C^{2}_{AC}$. Plotting these quantities for the current experimental setup of DUNE helps to understand the entanglement between $\nu_{\mu}$ and $\nu_{\tau}$ in a more rigorous way. 
For the present analysis the form of Eq. \ref{eq:2.12} modifies to,
\begin{equation}
\vspace{0cm}
\begin{split}
    C^{2}_{\mu e;\mu}+ C^{2}_{\mu\tau;\mu} \leq C^{\mu ^2}
\end{split}
\label{eq:4.6}
\end{equation}

\begin{figure}

\end{figure}

\begin{figure}  
\vspace{-1.5cm}
\includegraphics[scale=0.5]{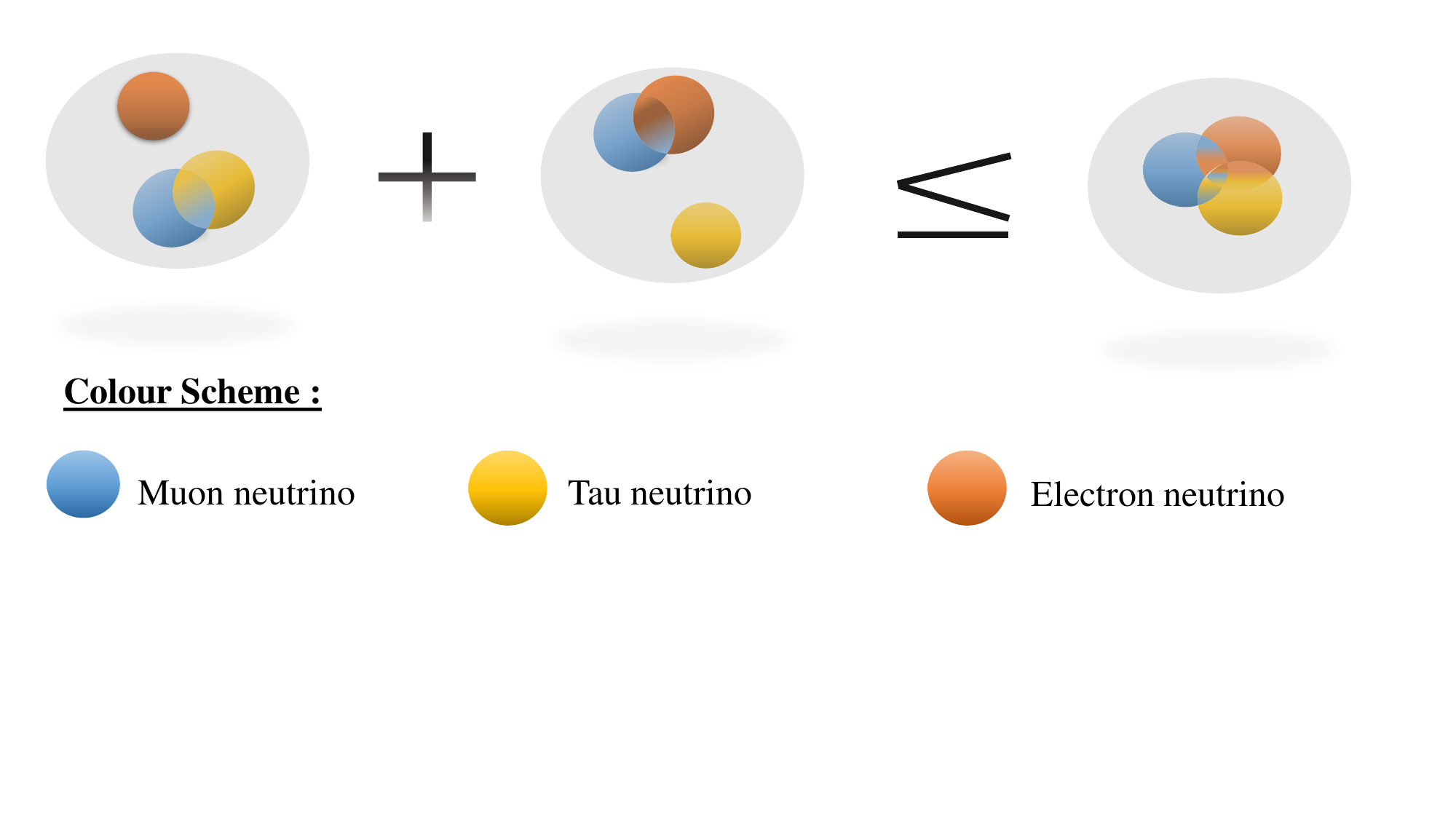}
    \vspace{-3.5cm}
    \\
    \includegraphics[scale=0.5]{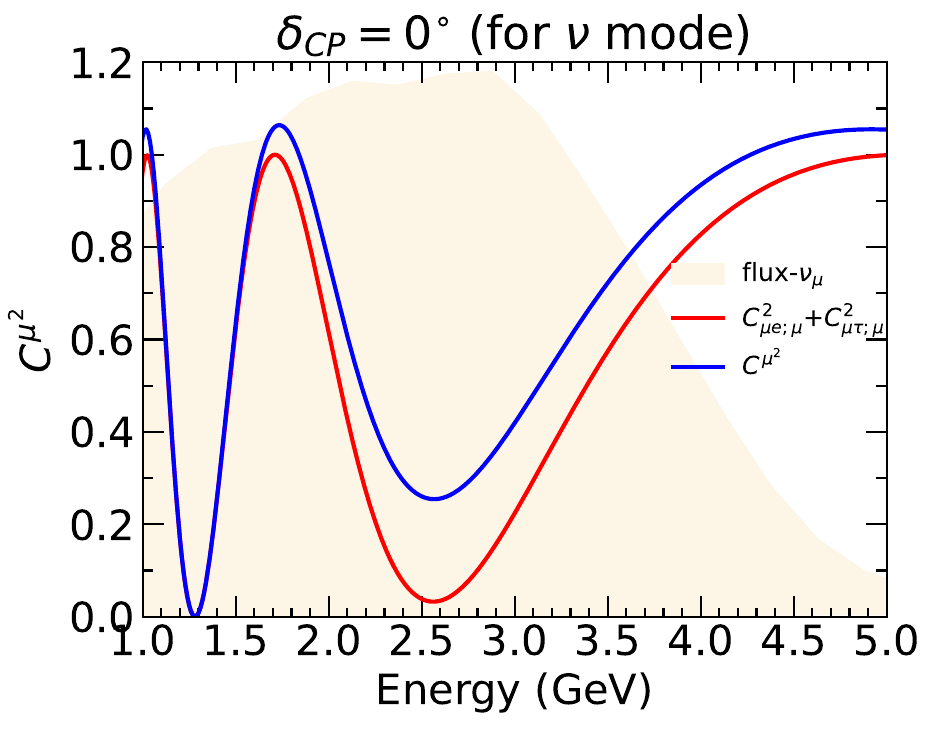}
\includegraphics[scale=0.5]{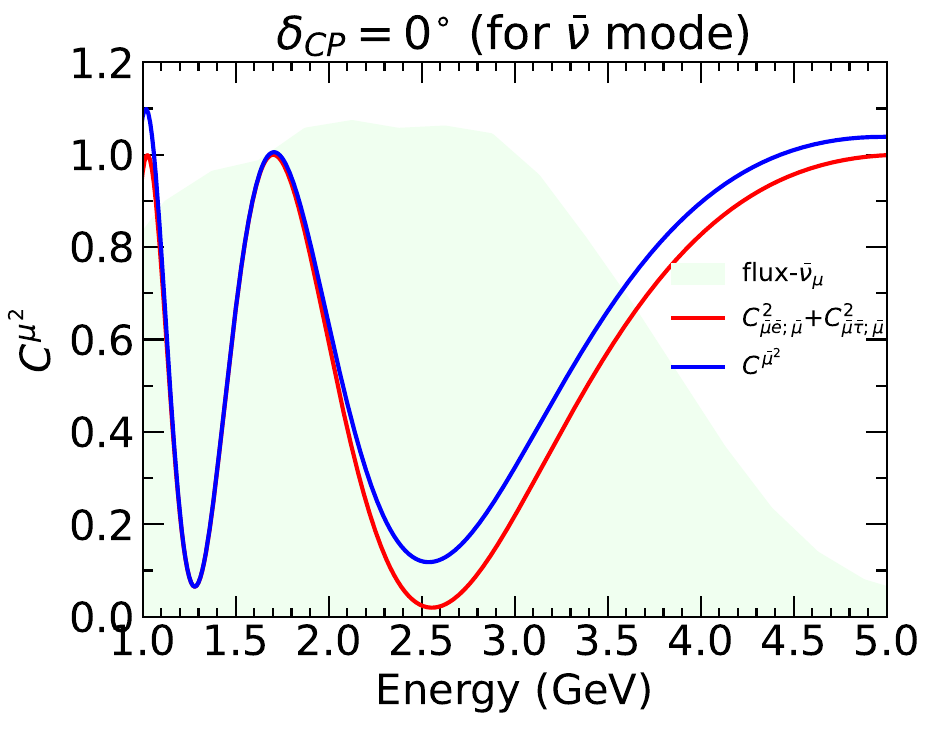}
    \caption{Upper panel shows Schematic picture of CKW-type monogamy relation. Left (right) panel of lower row shows the monogamy inequality in DUNE experimental configuration. Color codes are given in the legend.}
    \label{ckw-1}
\end{figure}
Schematic visualisation of the same is demonstrated by the upper panel of Fig. \ref{ckw-1}. In the lower panel of adjacent Fig. \ref{ckw-1}, the blue curve represents the total squared concurrence of the neutrino system, $C^{\mu^2}$, while the red curve corresponds to the term $ C_{\mu e;\mu}^2+ C_{\mu\tau;\mu}^2$. From the plot, it is clear that the three-flavor neutrino system satisfies the CKW inequality across all energy bins. Notably, within energy window EW1, the inequality approaches equality. This behavior can be attributed to the negligible contribution of the term $C_{\mu e \tau}^2 = 4P_{\mu e}P_{\mu\tau}$. As seen in Fig. \ref{fig:2}, $P_{\mu e}$ is significantly smaller than $P_{\mu\tau}$, indicating that $P_{\mu\tau}$ is the dominant factor in determining $C_{\mu e \tau}^2$. Interestingly, the inequality becomes valid when $P_{\mu e} \neq 0$ and $P_{\mu\tau} > 0.5$, which occurs within the energy windows EW2 and EW3. The above analysis is clear-cut evidence of the possibility of forming the W \cite{Yang_2004} state in  three-flavor neutrino.\\ From our analysis, it is evident that the $\nu_{e}$ flavor state acts as a separable state when the initial flavor state is $\nu_{\mu}$. Meanwhile, the $\nu_{\mu}$ and $\nu_{\tau}$ states share entanglement, leading to the formation of a W state within this experimental setup \cite{PhysRevA.72.022333}. The entangled W state for the neutrino system behaving as:
\begin{equation}
\vspace{-0.1cm}
    \ket{\psi(t)}=\Tilde{\mathbb{U}}_{\mu e}\big(\ket{1_{e}}\otimes\ket{0_{\mu}0_{\tau}}\big)+\ket{0_{e}}\otimes\big(\Tilde{\mathbb{U}}_{\mu\mu}\ket{1_{\mu}0_{\tau}}+\Tilde{\mathbb{U}}_{\mu\tau}\ket{0_{\mu}1_{\tau}}\big)
\label{eq:4.6}
\end{equation}
 In addition to the W state, the Greenberger-Horne-Zeilinger (GHZ) state, typically written as $\ket{000} + \ket{111}$, represents another fundamental class of tripartite entanglement. While both W and GHZ states are maximally entangled and cannot be interconverted via local operations, their entanglement structures differ markedly. In the GHZ state, a measurement on any qubit destroys entanglement among the remaining qubits, reflecting genuine three-way entanglement \cite{D_r_2000}. In contrast, the W state retains bipartite correlations even after the loss of a qubit, demonstrating its robustness and distributed nature \cite{Swain:2021btc, Sk:2021dbf}. This property, characterized by the Coffman-Kundu-Wootters (CKW) monogamy relation, suggests that in three-flavor neutrino oscillations, bipartite entanglement is expected to dominate over genuine tripartite correlations.


\subsection{Oscillation parameters at entanglement extrema}
In the previous subsections, we identified three distinct energy windows where EOF and concurrence exhibit ``global minimum", maximum, and ``local minimum" values. To summarize, EW1 shows $\nu_{\mu}$ in an almost pure state, being completely disentangled from $\nu_e$ and $\nu_{\tau}$, EW2 exhibits maximal entanglement among all three flavor states and in EW3, there is minimal mixing between $\nu_{\mu}$ and $\nu_{\tau}$, with $\nu_e$ remaining nearly separable. 
From an experimental perspective, precise measurements of neutrino oscillation parameters are most effectively conducted in energy windows where one of the three flavor states is either in a pure state or nearly disentangled. Specifically, EW1 provides a pure state for $\nu_{\mu}$, while EW3 offers a nearly separable state for $\nu_e$. In this subsection, we will focus on the energy points of 1.27 GeV and 2.5 GeV to determine the best-fit values for the oscillation parameters at these specific points. The main motivation to focus on these two points are following; at the concurrence minimum, the neutrino flavor state becomes nearly separable, minimizing the parameter uncertainty. Hence, oscillation parameters can be measured with the highest precision. In our analysis, both EOF and concurrence attain absolute minima at 1.27 GeV, where $P_{\mu\mu}$ approaches unity and the appearance probabilities of $\nu_e$ and $\nu_\tau$ vanish. This signifies a predominantly pure $\nu_\mu$ configuration, implying that oscillation parameters measured near this energy are expected to be more precise and reliable. Moreover, the ``local minimum'' point holds equal significance in our analysis for two main reasons. First, around 2.5 GeV, the entanglement reaches a minimum (though not exactly zero), indicating a nearly separable (but not fully separable) neutrino flavor state, which is particularly suitable for precise extraction of oscillation parameters. Second, the DUNE neutrino flux peaks at approximately 2.5 GeV, providing the highest event rate within this energy range and thereby improving the statistical precision of the results. To achieve this, we varied all five oscillation parameters ($\theta_{12}, \theta_{13}, \theta_{23}$ $,\Delta m_{21}^2$, and $\Delta m_{31}^2$) within their $3 \sigma$ confidence levels as specified in NuFit v6.0 \cite{Esteban:2024eli}, while considering the full range of $\delta_{CP}$ from $0^\circ$ to $360^\circ$. We performed a comprehensive analysis by scanning the oscillation parameters and identified the minimum concurrence value along with the associated parameters. This analysis was performed using the DUNE experimental configuration.

The results are summarized in Table \ref{bestfit}. The first row presents the best-fit values of oscillation parameters from NuFit v6.0. The second and third rows show the oscillation parameters at 2.5 GeV and 1.27 GeV, respectively, where the concurrence is at its minimum, with the minimum concurrence values noted in parentheses. Table \ref{bestfit} reveals notable differences between the oscillation parameters at these energy points and the values from NuFit v6.0. For instance, while $\theta_{12}$ is $33.68^\circ$ in NuFit v6.0, it shifts to $31.31^\circ$ and $35.71^\circ$ at 2.5 GeV and 1.27 GeV, respectively. Interestingly, $\theta_{13}$ values remain relatively consistent across these reference cases. However, the octant problem of the atmospheric angle shows that $\theta_{23}$, which is in the lower octant according to NuFit v6.0, shifts towards a maximally mixing state at both minimum concurrence points.
Additionally, solar and atmospheric mass squared differences also change from the NuFit values. Most notably, $\delta_{CP}$ trends towards the maximum CP-violating phase of $270^\circ$ when considering the minimum concurrence points. In the next section, we will explore how the concepts of CP violation, mass hierarchy, and the octant of the atmospheric mixing angle are influenced by the presence or absence of quantum entanglement.

\section{Results}
\label{resultss}
\begin{table}[]
    \begin{tabular}{||c||c||c||c||c||c||c||}
    \hline
    \hline
        Condition & $\theta_{12} (^{\circ})$  & $\theta_{13} (^{\circ})$  & $\theta_{23} (^{\circ})$  &  $\Delta m_{21}^2$(${\rm eV}^2$)  & $\Delta m_{31}^2$($ {\rm eV}^2$) & $\delta_{CP} (^{\circ})$ \\
        \hline
        \hline
        NuFit v6.0 \cite{Esteban:2024eli} &  $33.68^{\circ}$ & $8.56^{\circ}$  & $43.3^{\circ}$ 
 & $7.49 \times 10^{-5} $   & $2.513 \times 10^{-3}$ & $212^{\circ}$\\
 \hline
 \hline
 $@$ 2.50 GeV  (0.03225)& $31.31^{\circ}$  &  $8.43^{\circ}$  & $45.2^{\circ}$ 
 & $7.01 \times 10^{-5}$   & $2.487 \times 10^{-3} $ & $270^{\circ}$ \\
 \hline
 \hline
 $@$ 1.27 GeV  (0.00314)& $35.71^{\circ}$  &  $8.23^{\circ}$  & $45.2^{\circ}$ 
 & $7.61 \times 10^{-5}$   & $2.587 \times 10^{-3} $ & $270^{\circ}$ \\
 \hline
 \hline
    \end{tabular}
    \caption{First row presents the values of the oscillation parameters from NuFit v6.0, while the second (third) row shows the best-fit values obtained by considering the ``local (global) minima" of concurrence. The corresponding concurrence values are indicated in parentheses in the first column of the second and third rows. }
    \label{bestfit} 
\end{table}

\subsection{CP violation}  
CP violation sensitivity measures an experiment's ability to distinguish CP-violating values of $\delta_{CP}$ from CP-conserving values of $0^\circ$ and $180^\circ$. In this section, we will examine the CP violation sensitivity both in the presence and absence of quantum entanglement. To obtain these results, we varied the true value of $\delta_{CP}$ across the full range from $0^\circ$ to $360^\circ$, while keeping the other oscillation parameters fixed according to Table \ref{osc} in three different conditions. In the test spectrum, we used CP-conserving values of $\delta_{CP}$ (i.e., $0^\circ$ and $180^\circ$) and varied $\theta_{23}$ within its $3 \sigma$ confidence level, with the values of other oscillation parameters same as true values.

\begin{figure}
\includegraphics[scale=0.5]{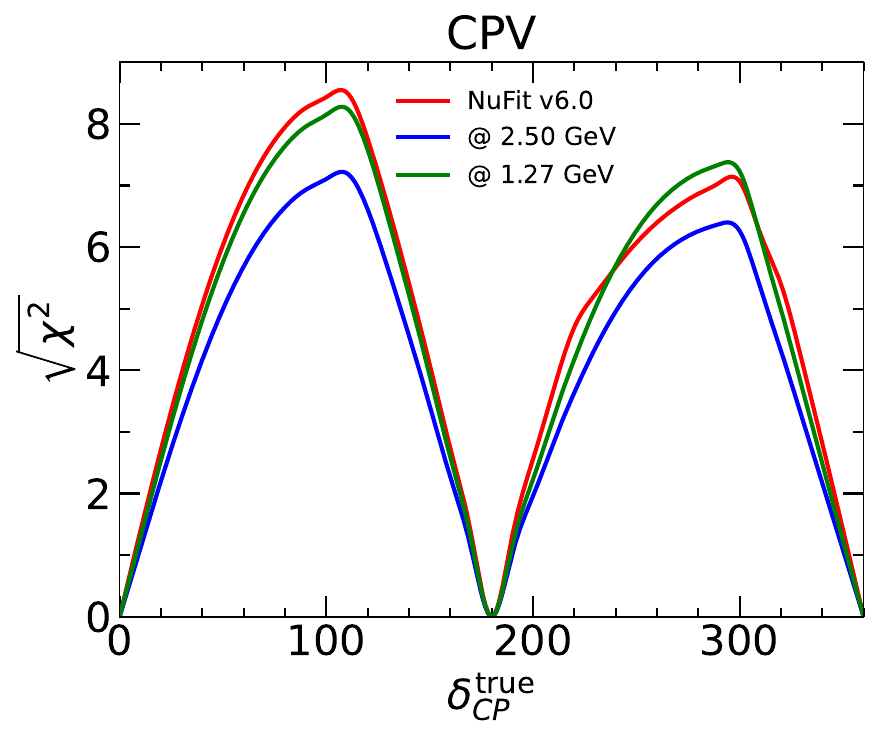}    \includegraphics[scale=0.5]{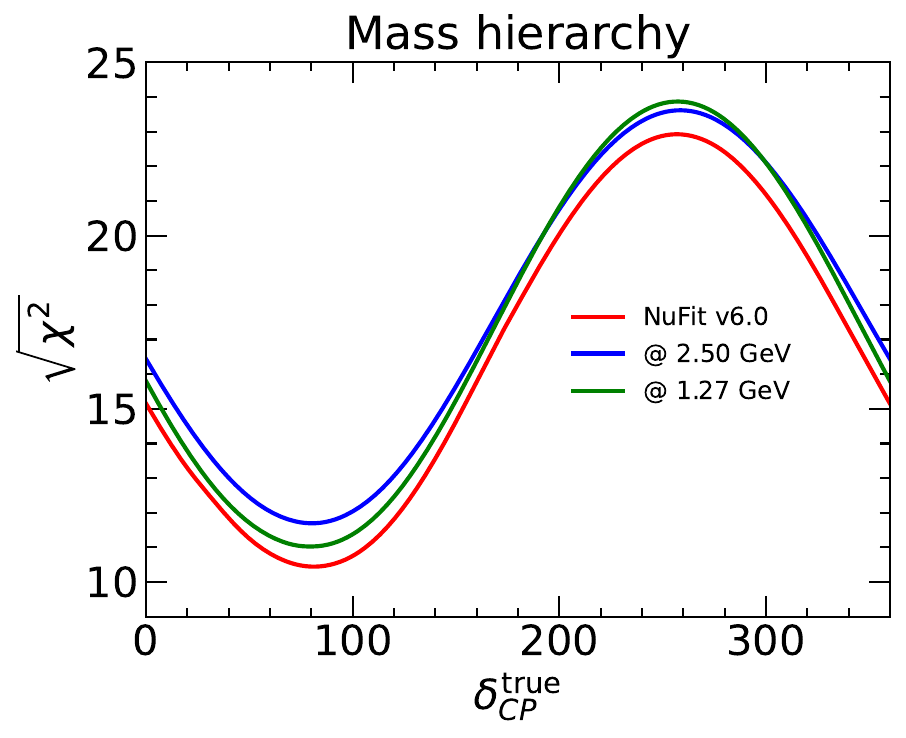}
    \caption{Physics sensitivities in the presence and absence of quantum entanglement concept. Left (right) panel is the CPV (mass hierarchy) sensitivity as a function of $\delta_{CP}^{\rm true}$. Color codes are given in the legend.}
    \label{result}
\end{figure}

The CP violation sensitivity results are illustrated in the left panel of Fig. \ref{result}. In this panel, the red, blue, and green curves represent the CP violation sensitivity as a function of the true value of $\delta_{CP}$ under three different conditions: the red curve corresponds to the true values of oscillation parameters from the first row of Table \ref{bestfit}, the blue and green curves represent the sensitivity using the best-fit values of oscillation parameters from the second and third rows of Table \ref{bestfit}, respectively.
In terms of quantum entanglement, if entanglement is not considered, the red curve shows the CP violation sensitivity for the DUNE configuration. When entanglement effects are included, the CP violation sensitivity decreases for both the ``global'' and ``local'' minimum concurrence points, reducing from $8.50\sigma$ to $8.25\sigma$ and $7.19\sigma$, respectively, for $\delta_{CP}^{\text{true}}$ values between $0^{\circ}$ and $180^{\circ}$. In the range $\delta_{CP}^{\text{true}} = [180^{\circ}–360^{\circ}]$, the highest CPV sensitivity corresponds to the ``global'' minimum concurrence point with a value of $7.33\sigma$, while the ``local'' and NuFIT data points yield maximum sensitivities of $6.37\sigma$ and $7.04\sigma$, respectively.

\subsection{Mass Hierarchy}
Mass hierarchy sensitivity measures an experiment's ability to differentiate between normal hierarchy (NH) and inverted hierarchy (IH). In this subsection, we will examine how mass hierarchy sensitivity is affected by quantum entanglement compared to the standard case. The right panel of Fig. \ref{result} displays the mass hierarchy sensitivity for the DUNE experimental setup. To generate these plots, we varied $\delta_{CP}^{\rm true}$ from $0^\circ$ to $360^\circ$, assuming NH for the true spectrum and using other oscillation parameters from Table \ref{bestfit} under three different conditions. For the test spectrum, we considered IH and varied $\delta_{CP}$ and $\theta_{23}$ within their full ranges and in $3 \sigma$ confidence level respectively.

In the plot, the red, blue, and green curves represent the mass hierarchy sensitivity in the absence of quantum entanglement, with ``local minima" of concurrence, and with ``global minima" of concurrence, respectively. The results show that the sensitivity is higher for the green and blue curves compared to the red curve, indicating that including the energy points where concurrence is minimized enhances mass hierarchy sensitivity. This holds true for both ``local" and ``global" minima of concurrence.

Octant sensitivity, which measures the ability to distinguish between the lower and upper octants of the atmospheric mixing angle, is assessed using the data from Table \ref{bestfit} for the true spectrum and the opposite octant for the test spectrum. Given that $\theta_{23}$ is nearly maximal ($45.2^\circ$) in the second and third rows of Table \ref{bestfit}, the octant sensitivity from these points is negligible. Consequently, we do not present a plot for octant sensitivity.

In summary, the concept of quantum entanglement significantly impacts experimental observations. Incorporating quantum entanglement alters the results for CP violation, mass hierarchy, and octant sensitivity, highlighting its importance in precision measurements.

\section{Concluding remarks}
In this study, we explored the effects of quantum entanglement within the framework of three-flavor neutrino oscillations, specifically in the context of the DUNE experiment. Our findings show the development of maximal entanglement between the muon neutrino ($\nu_{\mu}$) and tau neutrino ($\nu_{\tau}$), while the electron neutrino ($\nu_e$) remains nearly separable as a flavor state. To quantify the degree of entanglement, we use measurements such as EOF and concurrence, both of which clearly indicate the presence of entanglement in the three-flavor neutrino oscillation system. To further investigate whether this entanglement is bipartite or genuine tripartite, we apply the MOE inequality, specifically utilizing the CKW inequality. This inequality held true in the DUNE setup, showing that the system exhibits bipartite entanglement and suggesting the potential presence of W-state type entanglement.

Focusing on the behavior of EOF and concurrence, we identify three energy windows where ``global minima", maxima, and ``local minima" occur. For the ``local" and ``global" minima points, we derive the best-fit oscillation parameters by minimizing entanglement. We extend these results at the minima points to examine sensitivities to CP violation (CPV), mass hierarchy, and the octant of the atmospheric mixing angle. In the presence of quantum entanglement, CPV sensitivity was reduced from 
$8.50\sigma$ to $8.25\sigma$ and $7.19\sigma$, respectively, for $\delta_{CP}^{\text{true}}$ values between $0^{\circ}$ and $180^{\circ}$. In the range $\delta_{CP}^{\text{true}} = [180^{\circ}–360^{\circ}]$, the highest CPV sensitivity corresponds to the ``global'' minimum concurrence point with a value of $7.33\sigma$, while the ``local'' and NuFIT data points yield maximum sensitivities of $6.37\sigma$ and $7.04\sigma$, respectively. In contrast to CPV, the sensitivity to the mass hierarchy increase when considering the concept of entanglement. However, for octant sensitivity, due to the atmospheric mixing angle approaching maximal mixing at the points of minimum entanglement, there is no significant improvement in sensitivity in the presence of entanglement.

In conclusion, examining neutrino oscillations through quantum entanglement offers a clearer understanding of the quantum nature of the three-flavor neutrino system. For instance, within the DUNE configuration, we observe three significant energy windows, with the ``global minimum" at 1.27 GeV and a ``local minimum" at 2.5 GeV.
``Global minimum" corresponds to the fully disentangled $\nu_{\mu}$ state, while the ``local minimum" represents a nearly separable $\nu_e$ flavor state that maintains a non-zero entangled state with $\nu_{\mu}$ and $\nu_{\tau}$. From the viewpoint of quantum entanglement, the event rate at 1.27 GeV enables a more precise extraction of neutrino oscillation parameters compared to the event rate at 2.5 GeV. Therefore, to obtain more accurate information on the oscillation parameters in the DUNE experiment, it is crucial to focus on event rates at the 1.27 GeV energy window. Incorporating the concept of quantum entanglement into neutrino oscillation studies enhances the precision in measuring neutrino oscillation parameters.

\section*{Acknowledgments}
Rajrupa Banerjee would like to thank the Ministry of
Electronics and IT for the financial support through the
Visvesvaraya fellowship scheme for carrying out this research work.  RB is very thankful to Prof. Prasant K. Panigrahi for the
fruitful discussion carried from time to time for the betterment of this work.
PP wants to thank Prime Minister’s Research Fellows (PMRF) scheme for its
financial support.
RM would like to acknowledge University of Hyderabad IoE
project grant no. RC1-20-012. We gratefully acknowledge the use of CMSD HPC facility of
University of Hyderabad to carry out the computational works. SP would like to acknowledge the financial support under MTR/2023/000687 funded by SERB, Govt. of India.

\bibliographystyle{jhep}
\bibliography{entanglement_neutrino}
\end{document}